\documentclass{article}

\usepackage{arxiv}

\usepackage[utf8]{inputenc} % allow utf-8 input
\usepackage[T1]{fontenc}    % use 8-bit T1 fonts
\usepackage{hyperref}       % hyperlinks
\usepackage{url}            % simple URL typesetting
\usepackage{booktabs}       % professional-quality tables
\usepackage{amsfonts}       % blackboard math symbols
\usepackage{nicefrac}       % compact symbols for 1/2, etc.
\usepackage{microtype}      % microtypography
\usepackage{lipsum}		% Can be removed after putting your text content
\usepackage{graphicx}
\usepackage{natbib}
\usepackage{doi}
\usepackage{amsmath}

\title{Interval Valued Fuzzy Modeling and Indirect Adaptive Control of Quadrotor}

\author{ \href{https://orcid.org/0000-0003-0057-6432}{\includegraphics[scale=0.06]{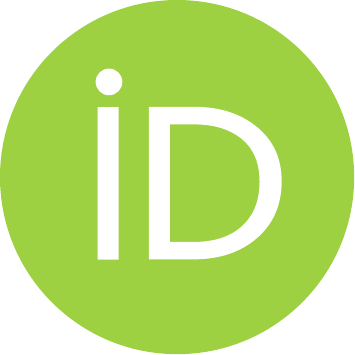}\hspace{1mm}Moufid ~Bouhentala}\thanks{Use footnote for providing further
		information about author (webpage, alternative
		address)---\emph{not} for acknowledging funding agencies.} \\
	LAAAS, Department of Electronics Engineering\\
	Mostefa Ben Boulaïd University\\
	Batna, Algeria \\
	\texttt{bouhentalam@gmail.com} \\
	\And
	\href{https://orcid.org/0000-0000-0000-0000}{\includegraphics[scale=0.06]{orcid.pdf}\hspace{1mm}Mouna ~Ghanai} \\
	Department of Electronics Engineering\\
	Mostefa Ben Boulaïd University\\
	\texttt{m.ghanai@univ-batna2.dz} \\
         \And
         \href{https://orcid.org/0000-0000-0000-0000}
   {\includegraphics[scale=0.06]{orcid.pdf}
        \hspace{1mm}Kheireddine ~Chafaa} \\
	Department of Electronics Engineering\\
	Mostefa Ben Boulaïd University\\
		\texttt{chafaak@gmail.com} \\
}

\hypersetup{
pdftitle={A template for the arxiv style},
pdfsubject={q-bio.NC, q-bio.QM},
pdfauthor={David S.~Hippocampus, Elias D.~Striatum},
pdfkeywords={First keyword, Second keyword, More},
}

\begin{document}
\maketitle

\begin{abstract}
%	\lipsum[1]
In this paper, a combination of fuzzy clustering estimation and sliding mode control is used to control a quadrotor system, whose mathematical model is complex and has unknown elements, including structure, parameters, and so on. In addition, they may be affected by external environmental disturbances. At first, the nonlinear unknown part of the system is estimated by a fuzzy model, A new method is presented for constructing a Takagi-Sugeno (TS) interval-valued fuzzy model (IVFM) based on input-output data of the identified system. Following the construction of the fuzzy model that estimates the unknown part of the quadrotor system, a control and on-line adjusting of the fuzzy modeled part of dynamics is used. In this step, the system model will be estimated in adaptive form so that the dynamic equations can be used in sliding mode control. Finally, the proposed technique is applied, and the simulation results are presented to show the effectiveness of this approach in controlling the quadrotor with unknown nonlinear dynamics.
 
\end{abstract}

% keywords can be removed

\section{Introduction}

The quadrotor unmanned aerial vehicle (UAV) is a typical second-order, nonlinear, strongly coupled dynamic system. An adaptive control scheme is significant for unmanned quadrotor helicopters, and this has received attention in the literature. External disturbances can have a impact on the stability and reliability of these systems. Therefore, it is very important to study the effects of system disturbances and design control strategies that address actuator faults and disturbances. To date, many methods and strategies have been presented in the literature \cite{[1]}. Authers presents PD control as well as a fuzzy adaptive PD control scheme in which the controller's gains are automatically adjusted in relation to the error signal rate \cite{[2]}. The fuzzy adaptive PD controller is quicker at achieving the desired response as compared to the other one. The authors of \cite{[3]} used a robust fuzzy control approach to stabilize a quadrotor's attitude angles. They used the parallel distributed compensation technique to design the fuzzy controller. High-order sliding mode observer was designed to provide state information required for fuzzy controller design.
The study \cite{[4]}, integrated integral backstepping with fuzzy logic to improve the flight controller's performance. Simulation results show superior performances for the fuzzy-integral backstepping approach over conventional integral backstepping. 
Fuzzy based sliding mode control for quadrotor UAV has recently been reported in \cite{[5]} and control structure is augmented via a genetic algorithm for optimization purposes. This approach can be further investigated using second order sliding mode control to avoid chattering. In \cite{[6]}, NRBF and E-RBF are used collectively to approximate the unknown dynamics of the quadrotor. A study \cite{[7]}, addresses online learning of the dynamics of a quadrotor nonlinear model through neural networks. Online learning is normally preferred over offline learning for the reason that quadrotors may fly in a dynamic environment where different worse case perturbations may occur, and such operating scenarios are hard to generate offline. Also, this study conveys the idea of using NNs to control the six degrees of freedom of a quadrotor along with designing an ANN observer to estimate states of the system. A neural network is used to approximate unmodeled dynamics during flight in \cite{[8]}, and interconnection errors in flight formation are investigated. in \cite{[9]} the position controller considered as nonlinear system and approximated using radial basis function neural networks, This study is unique in a sense that a complete controller is approximated via adaptive rule and trained online. This neuronal network-based controller provides adaptation for parametric uncertainties along with other external disturbances. 
In \cite{[10]} authors proposed a nonlinear-sliding-surface-based SMC for tracking and stabilization with an SMO. To improve the performance of the quadrotor system, they remodeled the augmented quadrotor dynamics and estimated the unmeasured states and their derivatives through the sliding mode observer. To improve the SMC control performance, a nonlinear sliding surface is used to reduce the convergence time with a modified deceleration curve. The authors proved the performance of a tracking system.
Based on the above research analysis, we propose an adaptive control strategy, and design an adaptive controller for a quadrotor UAVs in the presence of external and internal disturbances. The introduced adaptive control strategy is based on fuzzy logic control theory and adaptive sliding mode theory. A new method for constructing a Takagi-Sugeno (TS) type-2 fuzzy model offline, based on input-output data, is proposed, and the obtained fuzzy model of the dynamic is used to build the model-based control of the system. with an adaptive control strategy scheme that compensates the effects of the disturbance and an unmodeled dynamics control system. The system’s global asymptotic stability is validated by the Lyapunov function. The effectiveness and feasibility of the proposed method are demonstrated by simulation studies of the unmanned vehicle.
The rest of the paper is organized as follows. Section ~\ref{fuzzy_modeling} describes the proposed method of IVFM construction from input-output datasets of the system. Section ~\ref{fuzz_quad_model} presents the application of the proposed method to build the IVFM of the quadrotor. Section ~\ref{ind_ctrl} presents the proposed control strategy based on fuzzy logic control and adaptive sliding mode for the quadrotor UAV under external disturbances and parameter changes. The stability analysis of the new methodology is validated by the Lyapunov function. The results of simulation studies that demonstrate the performance of the proposed adaptive fuzzy schemes are presented in Section ~\ref{res_discu}. Finally, general conclusions and future work plans are provided.

\section{Fuzzy modeling }
\label{fuzzy_modeling}

The proposed method for constructing a Takagi-Sugeno (TS) type-2 fuzzy model, based on the input-output data of the identified system is constructed in three steps: (1) structure identification by fuzzy clustering; (2) envelope detection; and (3) parameter identification. In the structure identification phase, a clustering method based on the Gustafson-Kessel algorithm (GKCA) is used in order to detect the linear subsystems of the whole nonlinear system \cite{[11]} (local linearization). Then, an envelope detection algorithm (EDA) based on the derivative concept is proposed to estimate both the upper and lower membership functions of the type-2 fuzzy membership function (T2MF) defined point-wise. In the parameter identification step, the least squares algorithm is applied to compute the best parameter values for the premises (Gaussians) and the consequences (straight lines) parameters.

\subsection{Fuzzy Clustering overview}

Clustering is the partitioning of data into subsets or groups based on similarities between the data. The main potential of clustering is to detect the underlying structure in data, not only for classification and pattern recognition, but for model reduction and optimization. To detect clusters of different geometrical shapes in a single data set, we used the Fuzzy Gustafson Kessel Clustering (FGK), which extends the standard fuzzy c-means algorithm by employing an adaptive distance norm \cite{[12],[13],[14]} . Each cluster has its own norm-inducing matrix, $A_i$, which yields the inner-product norm Eqn.~(\ref{norm}) shown below:

\begin{equation}
D_{ikA}^2=({\mathbf{x}}_k-{\mathbf{v}}_i)^T{A_i}({\mathbf{x}_k}-\mathbf{v}_i),
\label{norm}
\end{equation}
where $1 \leq i\leq c, 1 \leq k \leq N$ .
The matrices $A_i$  are used as optimization variables in the $C-means$ functional, which will allow to each cluster to adapt the distance norm to the local topological structure of the data. Let $A$ denote a c-tuple of the norm-inducing matrices: $A=({A_1},{A_2},...,{A_c})$.  The objective functional of the GK algorithm is defined by:

\begin{equation}
J(\mathbf{X},\mathbf{U\;,V,A}) = \sum\limits_{i = 1}^c \sum\limits_{k = 1}^N {(\mu _{ik})}^m\textnormal{D}_{ikA}^2 
\end{equation}
 
where $N$  the number of data points, $c$ the number of clusters, $x_k$  the $k^{th}$  data point, ${v_i}$  the $i^{th}$  cluster center, ${\mu _{ik}}$ the degree of membership of the $k^{th}$ data in the $i^{th}$ cluster, $A_i$  the norm-inducing matrix of the $i^{th}$ cluster and $m$ is a weighting exponent that determines the fuzziness of the resulting clusters (typically $m = 2$). 
In our paper, the GKCA is applied in order to obtain the fuzzy partition matrix $U ={[\mu _{ik}]_{c \times N}} $, with ${\mu _{ik}} \in [ {0\,\,1}]$  a membership degree.

\subsection{Proposed envelope detection and estimation algorithm}

The individual partitioning subsets  ${\mu _{ik}}$ projected on the regressor $X$ will define directly fuzzy regions, in which the data can be reasonably approximated by linear sub-models defined by clusters. Projection data can also make the fuzzy system easy to read by humans

\subsubsection{Detection of lower and upper premise MF} 

\begin{figure} 
%\centerline{\includegraphics[width=4.0in]{figure/fig1.png}}
\centering
\scalebox{0.6}{\includegraphics{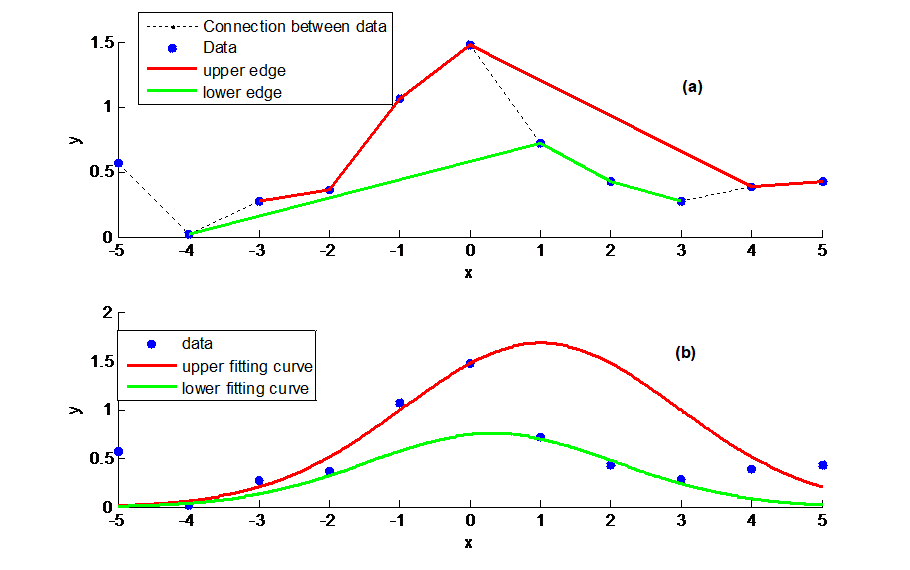}}
\caption{(a) Data envelope detection (b) Gaussian fitting}
\label{fig1}
\end{figure}

In order to detect the fuzzy premise MFs (lower MF $\underline{f}$ and  the upper MF $\bar f$) we propose in what follows an approach called for Envelope Detection Algorithm (EDA) based on derivative concept. The algorithm is described in the following algorithm: 

1)	First, we interpolate the points linearly (dotted curve in Fig.~\ref{fig1} (a). between all segments $[y(i),y({i+1})]$ where  $i$ represents the data indices.

The third-level of heading follows the style of the second-level heading.
2)	We compute the slops between each successive data points $[y(i),y(i + 1)]$ as:  

\begin{equation}
	d\left({i + 1} \right) = \frac{y\left({i + 1} \right) - y\left( i \right)}{x\left( {i + 1} \right) - x\left( i \right)}
\end{equation}
3) Two cases can arise:\\

\begin{itemize}
	\item When the slop $d$ in interval $[x(i), x(i+1)]$ is positive, then the point   is considered as higher membership point and memorized with its indices as $(I_h , y_h(I_h))$, where $I_h =i+1$ represents the indices of the corresponding higher membership point $y_h$ s.t. $y_h(I_h)=y(i+1)$. \\ 
	\item Otherwise, when the slop $d$ in interval $[x(i), x(i+1)]$ is negative, then the point   is considered as lower membership point and memorized with its indices as $(I_L , y_L(I_L))$, where $I_L =i+1$ represents the indices of the corresponding lower membership point $y_L$ s.t. $y_L(I_L)=y(i+1)$.  
\end{itemize}

By applying our method cited above, we show in Fig.~\ref{fig1} the shape of the obtained upper and lower envelopes, which represent the upper and lower type-2 MFs. The details of this technique are given in EDA Algorithm \cite{[15]}.
This algorithm doesn’t ensure the separation of envelopes; to do that, we can apply the same algorithm again for both the upper and lower sets. When we apply this algorithm to the initial set of points, we get two sets:$\underline{f}$ and $\overline{f}$. Next, we use the same algorithm for both sets $(\underline{f} and \overline{f})$   to get another four sets. The proposed algorithm can be presented as the algorithm described here :

\begin{enumerate}
	\item Applying ADE(initial set) gives two sets  $(\underline{f} , \overline{f})$  
	\item make $ i=j=1 $
	\item $(\underline{f}_j,{\bar f}_i)=ADE($ {initial set} $)$
	\item Calculate the upper set ${\bar f}$
	
	 \begin{enumerate}%[label=(\alph*)]
		 \item Make  $i = i+1 $
		 \item $({\underline{f}}_i,{\bar f}_i) = ADE(\bar f)$ 
		 \item If    go to step (a) else next:
		 \item  $\bar f = {\bar f_i}$
	 \end{enumerate}
	
  \item Calculate the lower set 	
   \begin{enumerate}%[label=(\alph*)]
	   \item Make  $j = j+1 $
	   \item $({\underline{f}}_j,{\bar f}_j) = ADE(\underline f)$ 
		 \item If  $j \ne j_{\max } $ go to step $(a)$ else next:
		 \item $\underline{f}  = {\underline{f}_j}$	  

   \end{enumerate}
\end{enumerate}

%%%%%%%%%%%%%%%%%%%%%%%%%%%%%%%%%%%%%%%%

%%%%%%%%%%%%%%%%%%%%%%%%%%%%%%%%%%%%%%%%%%%%%%%%%%%%%%%%%%%%%%%%%%%%%%

%
%\begin{figure}
%\begin{subfigure}{0.5\textwidth}
%\includegraphics[width=0.9\linewidth, height=6cm]{figure/fig2a.png}}
%\caption{(a) with first applying EDA one time,(b) envelope fitting with Gaussian model}
%\label{fig2a}
%\end{subfigure}
%\begin{subfigure}{0.5\textwidth}
%\includegraphics[width=0.9\linewidth, height=6cm]{figure/fig2b.png}}
%\caption{(c) after applying EDA two times, there are less intersection points (d) after Gaussian fitting, there is no intersection. }
%\label{fig2b}
%\end{subfigure}
%\end{figure}
%
%

%%%%%%%%%%%%%%%% begin figure %%%%%%%%%%%%%%%%%%%
%%% 3.34in is the maximum width you can have for a figure
\begin{figure} 
\centering
%\centerline{\includegraphics[width=4.0in]{figure/fig2a.png}}
\scalebox{0.6}{\includegraphics{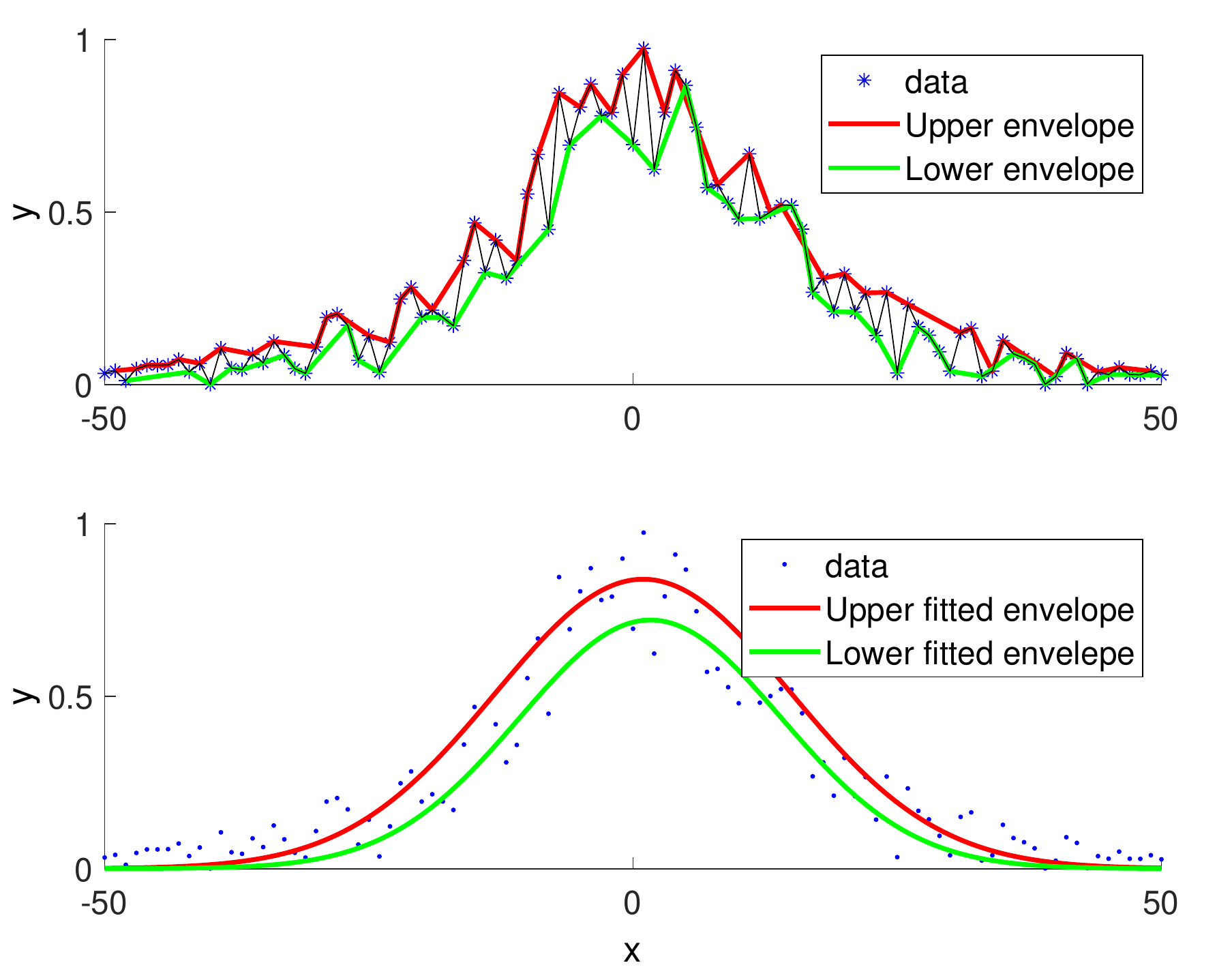}}
\caption{(a) with first applying EDA one time,(b) envelope fitting with Gaussian model}
\label{fig2a}
\end{figure}
%%%%%%%%%%%%%%%% end figure %%%%%%%%%%%%%%%%%%%
%%%%%%%%%%%%%%%%%%%%%%%%%%%%%%%%%%%%%%%%%%%%%%%%%%%%%%%%%%%%%%%%%%%%%%

\begin{figure} %[t]
%\centerline{\includegraphics[width=4.0in]{figure/fig2b.png}}
\centering
\scalebox{0.55}{\includegraphics{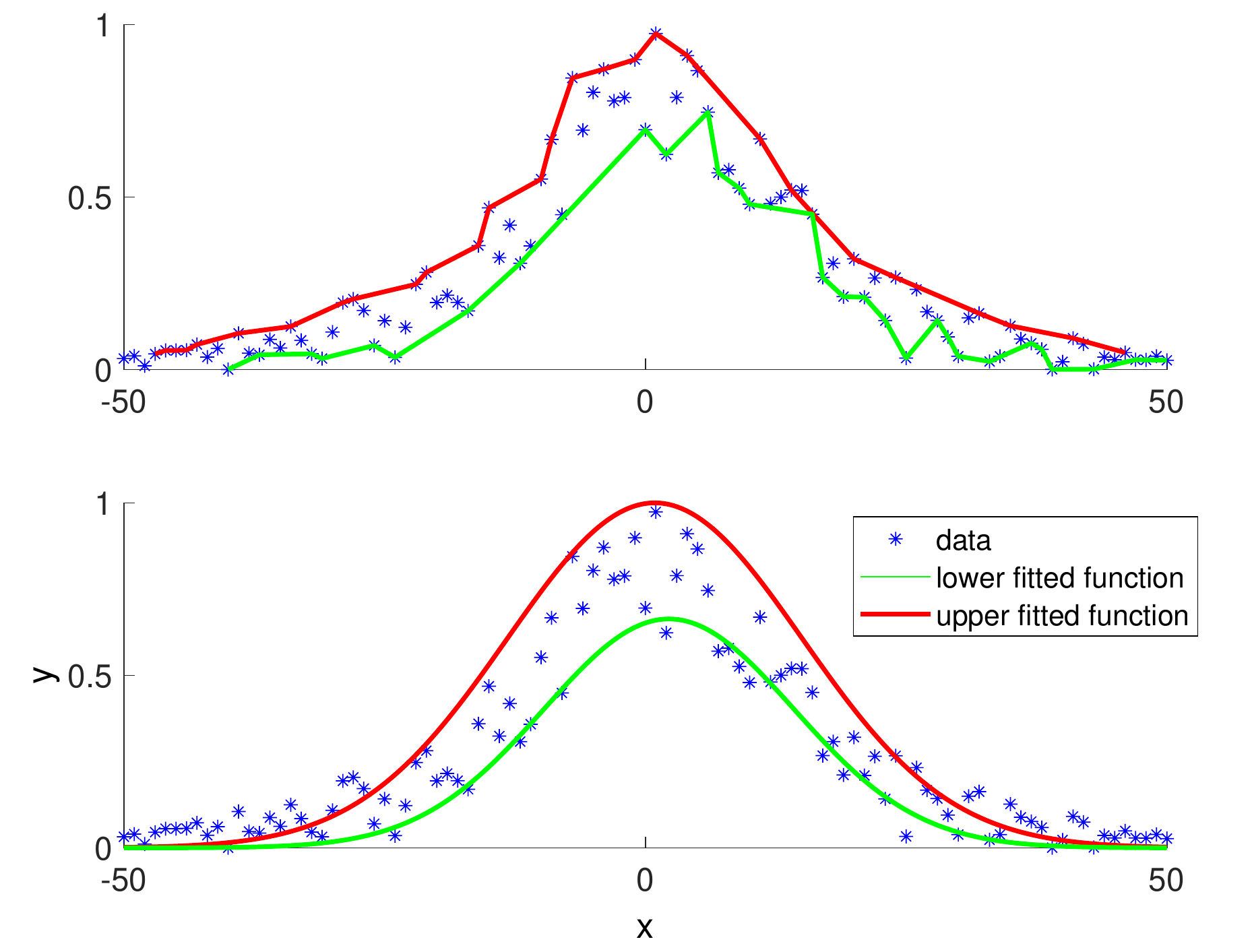}}
\caption{(c) after applying EDA two times, there are less intersection points (d) after Gaussian fitting, there is no intersection. }
\label{fig2b}
\end{figure}
%%%%%%%%%%%%%%%% end figure %%%%%%%%%%%%%%%%%%%
%%%%%%%%%%%%%%%%%%%%%%%%%%%%%%%%%%%%%%%%%%%%%%%%%%%%%%%%%%%%%%%%%%%%%%

Look to the next example where the initial function is a Gaussian with added noise to create a cloud to its shape, in the Fig.~\ref{fig2a} where the first ADE is applied and its corresponding fitting result as shown in the figure (b) there are many interactions points in different regions while the applying of EDA again make the envelopes less interacted and its fitted part show that our the final fitted functions envelops much points of the initial cloud Fig.~\ref{fig2b} (d).

%%%%%%%%%%%%%%%%%%%%%%%%%%%%%%%%%%%%%%%%%%%%%%%%%%%%%%%%%%%%%%%%%%%%%%
\subsection{Constructing fuzzy models from partitions}

 
At this step, we have two data sets of points: data set 1 represents the data of  $f_L$ and data set 2 represents the data of $f_U$ . Gaussian membership functions will be used in this investigation for the premise MF; thus, data sets 1 and 2 will be fitted to Gaussian models, yielding the lower and upper Gaussian membership functions $\bar f $ and $\underline f $.
Until now, we have obtained, by using the proposed method, type-2 fuzzy Gaussian membership functions for the premises. The global model can be conveniently represented as a set of affine Takagi-Sugeno (TS) rules $R_i$  as follows:

\begin{equation} 
\normalfont{if} x  \normalfont{ is } {\tilde A_i}  \normalfont{ then } \,{y_i} = {a_i}x + {b_i}\, \normalfont{for } i = 1,2,...,c
\label{rule}
\end{equation}

where $x$ is the input, and $y$ is the output, $ {\tilde A_i} = [\underline{f_i},\bar{f_i}] $   the $i^{th}$ obtained antecedent type-2 fuzzy MF. The consequent parameters $a_i$ and $b_i$ are estimated from the data using least square method.

\section{Fuzzy modeling of unknown dynamics of quadrotor }
\label{fuzz_quad_model}
In the following application, we will use the method presented above to model the unknown functions using input-output data, and then we will use those initial models to build the adaptive control.

\subsection{Dynamic model of Quadrotor}
Let’s consider the system in Eqn.~(\ref{sys}) as presented in \cite{[10],[16]} in which the state representation $\underline{x} = (\varphi ,\dot \varphi ,\theta ,\dot \theta ,\psi ,\dot \psi ,x,\dot x,y,\dot y,z,\dot z)$ is the state vector and  $U = \left[ {{U_\phi}\,{U_\theta}\,{U_\psi}\,{U_z}} \right]$ is the control vector. The transformation matrix between the rate of change of the orientation angles $(\dot \varphi ,\dot \theta ,\dot \psi )$  and the body angular velocities  $\left( {p,q,r} \right)$ can be considered as unity matrix if the perturbations from hover flight are small. Then, one can write $(\dot \varphi ,\dot \theta ,\dot \psi ) \approx \left( {p,q,r} \right)$  \cite{[16]}. Each axis can be approximately decoupled and controlled independently when the angular velocities are low and the attitude angle changes in small range. 
The drone parameters are shown in the Table~\ref{table_quad}.
  %= \left( {\begin{array}{*{20}{c}}
\begin{equation}
  {\dot x = \left(	\begin{array}{*{20}{c}}
  {\dot \phi } \\ 
  {\dot \theta \dot \psi {a_1} + \dot \theta {a_2}{\Omega _r} + {b_1}{U_1}} \\ 
  {\dot \theta } \\ 
  {\dot \phi \dot \psi {a_3} + \dot \phi {a_4}{\Omega _r} + {b_2}{U_2}} \\ 
  {\dot \psi } \\ 
  {\dot \theta \dot \phi {a_5} + {b_3}{U_3}} \\ 
  {\dot z} \\ 
  {g - (\cos \phi \cos \theta )\frac{U_z}{m}} \\ 
  {\dot x} \\ 
  {{u_x}\tfrac{U_z}{m}} \\ 
  {\dot y} \\ 
  {{u_y}\tfrac{U_z}{m}} 
\end{array}\right) \\ }
\label{sys}
\end{equation}

\[\left. \begin{array}{*{20}{c}}
  {a_1} = ({I_{yy}} - {I_{zz}})/{I_{xx}} \\
  {a_2} = {J_r}/{I_{xx}} \hfill \\
  {a_3} = ({I_{zz}} - {I_{xx}})/{I_{yy}} \\
  {a_4} = {J_r}/{I_{yy}} \hfill \\
  {a_5} = ({I_{xx}} - {I_{yy}})/{I_{zz}} \\ 
\end{array}  
\right|\begin{array}{*{20}{c}}
  b_1 = l/I_{xx}\\
  b_2 = l/I_{yy}\\ 
  b_3 = 1/{I_{zz}} \\ 
  u_x = (\operatorname{c} \phi \operatorname{s} \theta \operatorname{c} \psi  - \operatorname{s} \phi \operatorname{s} \psi ) \\ 
  u_y = (\operatorname{c} \phi \operatorname{s} \theta \operatorname{s} \psi  - \operatorname{s} \phi \operatorname{c} \psi ) 
\end{array}\]

%%%%%%%%%%%%%%% begin table   %%%%%%%%%%%%%%%%%%%%%%%%%%

%%%%%%%%%%%%%%%
\begin{table}[t]
\caption{Parameters of the Quadrotor}
\begin{center}
\label{table_quad}
%\resizebox{\columnwidth}{!}{%
\begin{tabular}{c l l}
%& & \\ % put some space after the caption
%\hline
%Example & Time & Cost \\
\hline
$I_{zz} = 0.09 kg.m^2 $& $m = 1.2 kg $    & $l=  0.23m $      \\
$I_{yy} =0.18  kg.m^2 $ & $d =1.1e-6 N.m.s^2  $ & $g =9.81m.s^-2 $\\
$I_{xx} =0.18 kg.m^2 $ & $b =54.2e-6 N.m^2 $ &  $j_r =1.32e^-3kg.m^2 $ \\       
 
\hline
\end{tabular}
%}
\end{center}
\end{table}

%%%%%%%%%%%%%%%% end table %%%%%%%%%%%%%%%%%%% 

the moment and force caused by the propeller are proportional to the square of the angular
velocity. The inputs of each axis can be composed of the speed of the propeller and described as \cite{[16]}:

\[\left. \begin{array}{*{20}{c}}
U_\phi = (\Omega_1 -\Omega_3 )\\
U_\theta = (\Omega_4 -\Omega_2 )\\
\end{array}  
\right|\begin{array}{*{20}{c}}
U_\phi = (\Omega_1 +\Omega_3 -\Omega_2 -\Omega_4 )\\
U_z = (\Omega_1 +\Omega_2 +\Omega_3 +\Omega_4 )\\
\end{array}\]

%%%%%%%%%%%%%%%%%%%%%%%%%%%%%%%%%%%%%%%%%%%%%%%%%%%%%
\subsection{Interval values Fuzzy modeling }
\label{IVFM_quad}
The model Eqn.~(\ref{sys}) can be decomposed into three subsystems:

\begin{equation}
 \left({\begin{array}{*{20}{c}}
  {{{\dot x}_1}} \\ 
  {{{\dot x}_2}} 
\end{array}} \right) = \left( {\begin{array}{*{20}{c}}
  {\dot \phi } \\ 
  {{f_\phi}(\dot \theta ,\dot \psi ) + {g_1}{U_1}} 
\end{array}} \right)
\label{sub1}
\end{equation}

\begin{equation}
\left( {\begin{array}{*{20}{c}}
  {{{\dot x}_3}} \\ 
  {{{\dot x}_4}} 
\end{array}} \right) = \left( {\begin{array}{*{20}{c}}
  {\dot \theta } \\ 
  {{f_\theta} + {g_2}{U_2}} 
\end{array}} \right)
\label{sub2}
\end{equation}

\begin{equation}
\left( {\begin{array}{*{20}{c}}
  {{{\dot x}_5}} \\ 
  {{{\dot x}_6}} 
\end{array}} \right) = \left( {\begin{array}{*{20}{c}}
  {\dot \psi } \\ 
  {{f_\psi} + {g_3}{U_3}} 
\end{array}} \right)
\label{sub3}
\end{equation}

where $f_{\phi}(\dot \theta ,\dot \psi ,{\Omega _r}) = \dot \phi \dot \psi {a_1} + \dot \phi {a_2}{\Omega _r}$ ,$f_{\theta}(\dot \theta ,\dot \psi ,{\Omega _r}) = \dot \theta \dot \phi {a_3} + \dot \theta {a_4}{\Omega _r}$ ,$f_{\psi}(\dot \theta ,\dot \psi ,{\Omega _r}) = \dot \theta \dot \psi {a_5}$ and ${g_\phi} = {b_1}$  ,${g_\theta} = {b_2}$ , ${g_\psi} = {b_3}$. The parameters $g_1$, $g_2$ ,$g_3$ and the functions $f_{\phi}$, $f_{\theta}$ and $f_{\psi}$ are considered as unknown functions, and they will be estimated as follow :

the first step is to identify the functions  ${f_i}(x ,\dot \psi ,{\Omega _r})$ and ${g_i}$ where $\left\{{i = \phi,\theta,\psi} \right\}$ using clustering. ${\hat f_i}$  can be written as an TS approximation of the form IVFM Eqn.~(\ref{rule}) :

\[
\begin{array}{*{20}{l}}
  {f_{\phi}(x) = {{\hat f}_{\phi}}({x_{8}},{x_{12}},{\Omega _r},{\theta _{f_{\phi}}}) + \varepsilon  = {\theta _{{f_{\phi}}}}^T{\varphi _{f_{\phi}}}({x})\;} \\ 
  {f_{\theta}(x) = {{\hat f}_{\theta}}({x_{10}},{x_{12}},{\Omega _r},{\theta _{f_{\theta}}}) + \varepsilon = {\theta _{{f_{\theta}}}}^T{\varphi _{{f_{\theta}}}}(x)} \\ 
  {f_{\psi}(x)= {{\hat f}_{\psi}}({x_{10}},{x_{12}},{\theta _{f_{\psi}}}) + \varepsilon = {\theta _{{f_{\psi}}}}^T{\varphi _{{f_{\psi}}}}(x)} 
\end{array}
\]

where $\theta _{f_i} = {\left[ {{\underline{w}}_l}^T,{{\underline{w}}_r}^T \right]_i}^T$ and ${\underline{\psi} } _{f_i}(x,\dot x) = \frac{1}{2}{\left[ {{{\underline{\xi } }_l}^T(x,\dot x),{{{\underline{\xi} } }_r}^T(x,\dot x)} \right]_i}^T$. \\

\begin{figure} [t]
\begin{center}
%\centerline{\includegraphics[width=4.0in]{figure/fig3.png}}
\scalebox{0.6}{\includegraphics{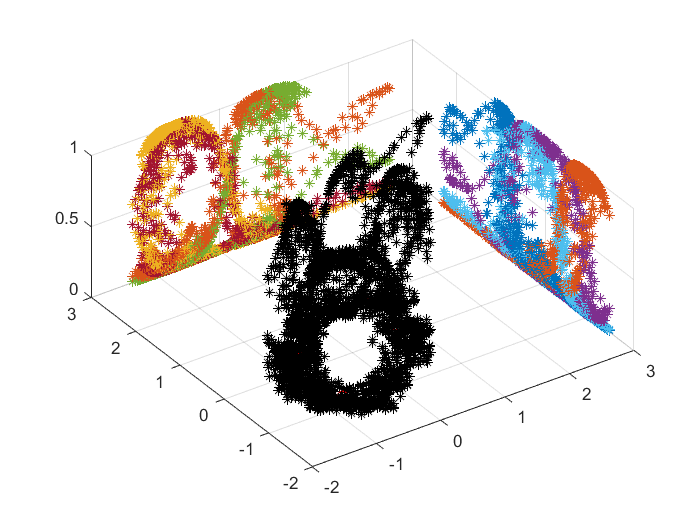}}
\caption{Projection and construction of membership functions }
\label{fig3}
\end{center}
\end{figure}

\begin{figure}
\begin{center}
\scalebox{0.6}{\includegraphics{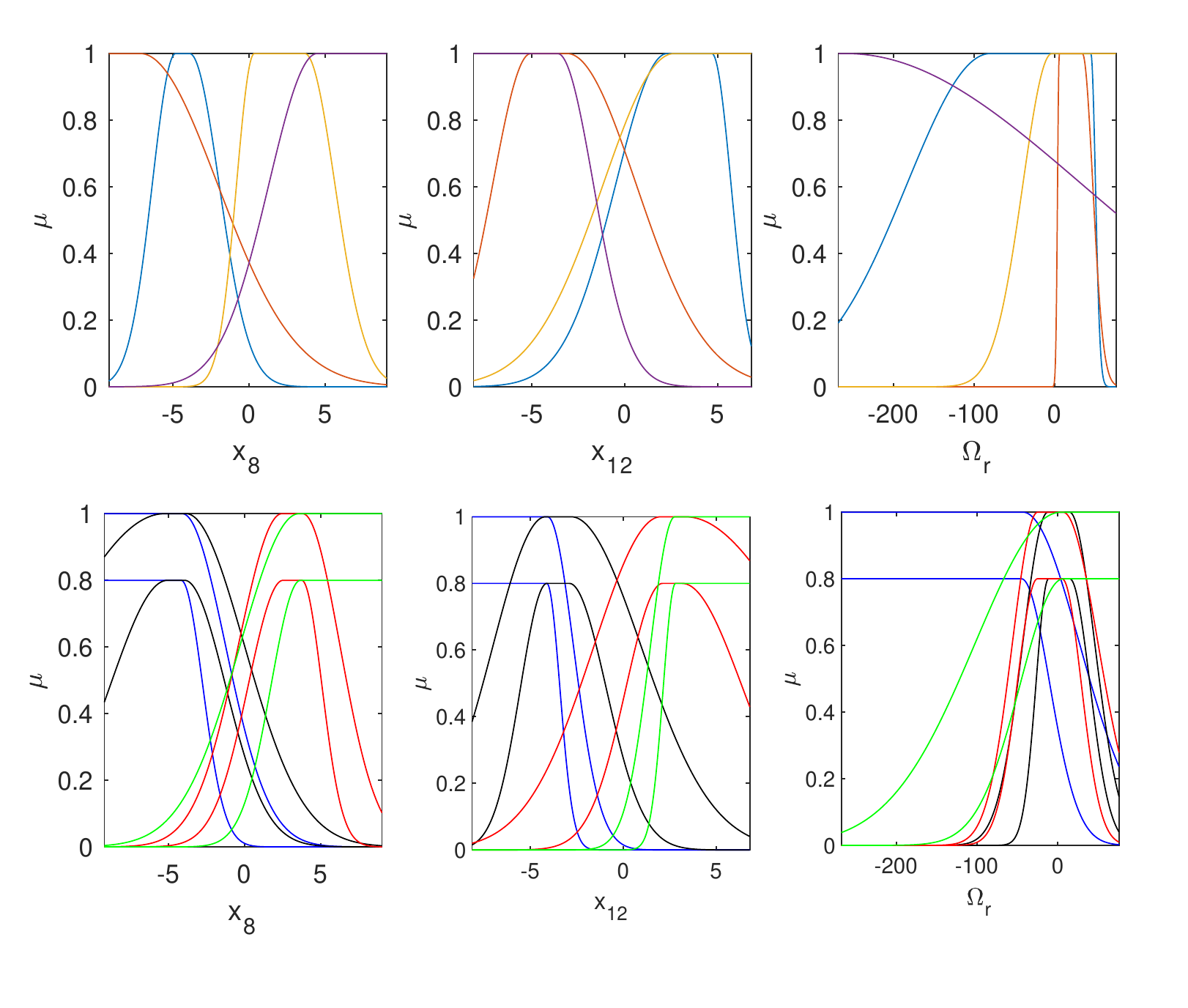}}
%\centerline{\includegraphics[width=4.0in]{figure/fig4.pdf}}
\caption{Membership Functions of ${\hat f_{\phi}}(\underline{x})$ }
\label{fig4}
\end{center}
\end{figure}

To collect input-output data, we generate two sets of data, one for identification by clustering, and the second for  validation. We start with a sequence of $4000$ samples ($2000$ for identification and $2000$ for validation) The sampling time is $T=0.01s$ with normal noise ($05\%$ of amplitude) is added to every output. The obtained data is treated using EDA in order to construct the membership functions as presented in Fig.~\ref{fig3}. The membership functions type-1 and type-2 obtained with simulation data for ${f_{\phi}}({x} ,\Omega ), {f_{\theta}}({x} ,\Omega )$  and ${f_{\psi}}({x} ,\Omega )$ are depicted in  The figures (Fig.~\ref{fig4}, ~\ref{fig6}), respectively. 

%%%%%%%%%%%%%%%% begin figure %%%%%%%%%%%%%%%%%%%
%%% 3.34in is the maximum width you can have for a figure

\begin{figure} 
\begin{center}
%\centerline{\includegraphics[width=4.0in]{figure/fig5.pdf}}
\scalebox{0.6}{\includegraphics{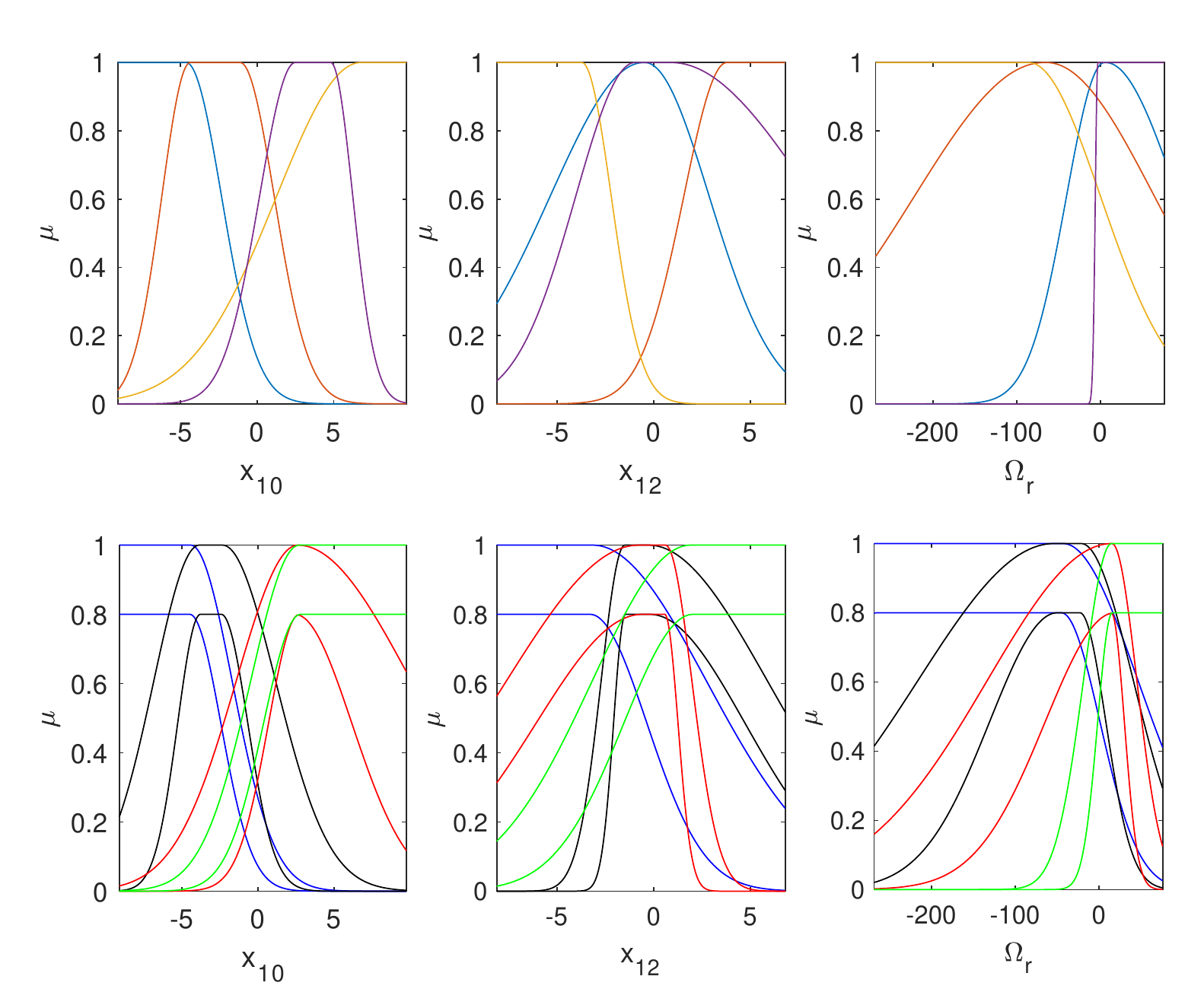}}
\caption{ Membership Functions of ${\hat f_{\theta}}(\underline{x})$}
\label{fig5}
\end{center}
\end{figure}

\begin{figure}
\begin{center}
%\centerline{\includegraphics[width=4.0in]{figure/fig6.pdf}}
\scalebox{0.6}{\includegraphics{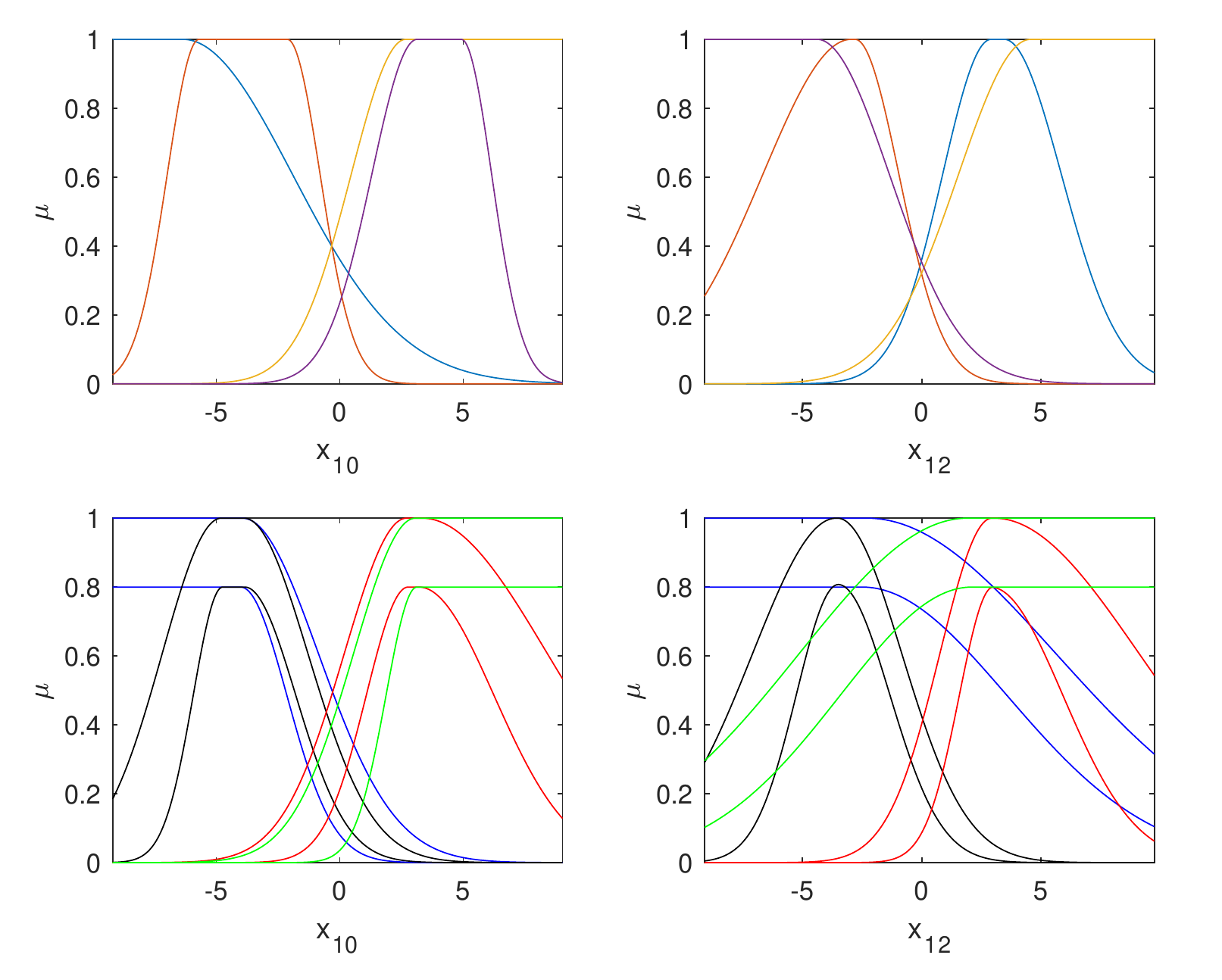}}
\caption{Membership Functions of ${\hat f_{\psi}}(\underline{x})$ }
\label{fig6}
\end{center}
\end{figure}

%\scalebox{0.5}{\includegraphics{union.pdf}}
%%%%%%%%%%%%%%%%%%%%%%%%%%%%%%%%%%%%%%%%%%%%%%%%%%%%%

\begin{figure}[!htb]
\begin{center}
%\centerline{\includegraphics[width=4.00in]{figure/fig7.pdf}}
\scalebox{0.6}{\includegraphics{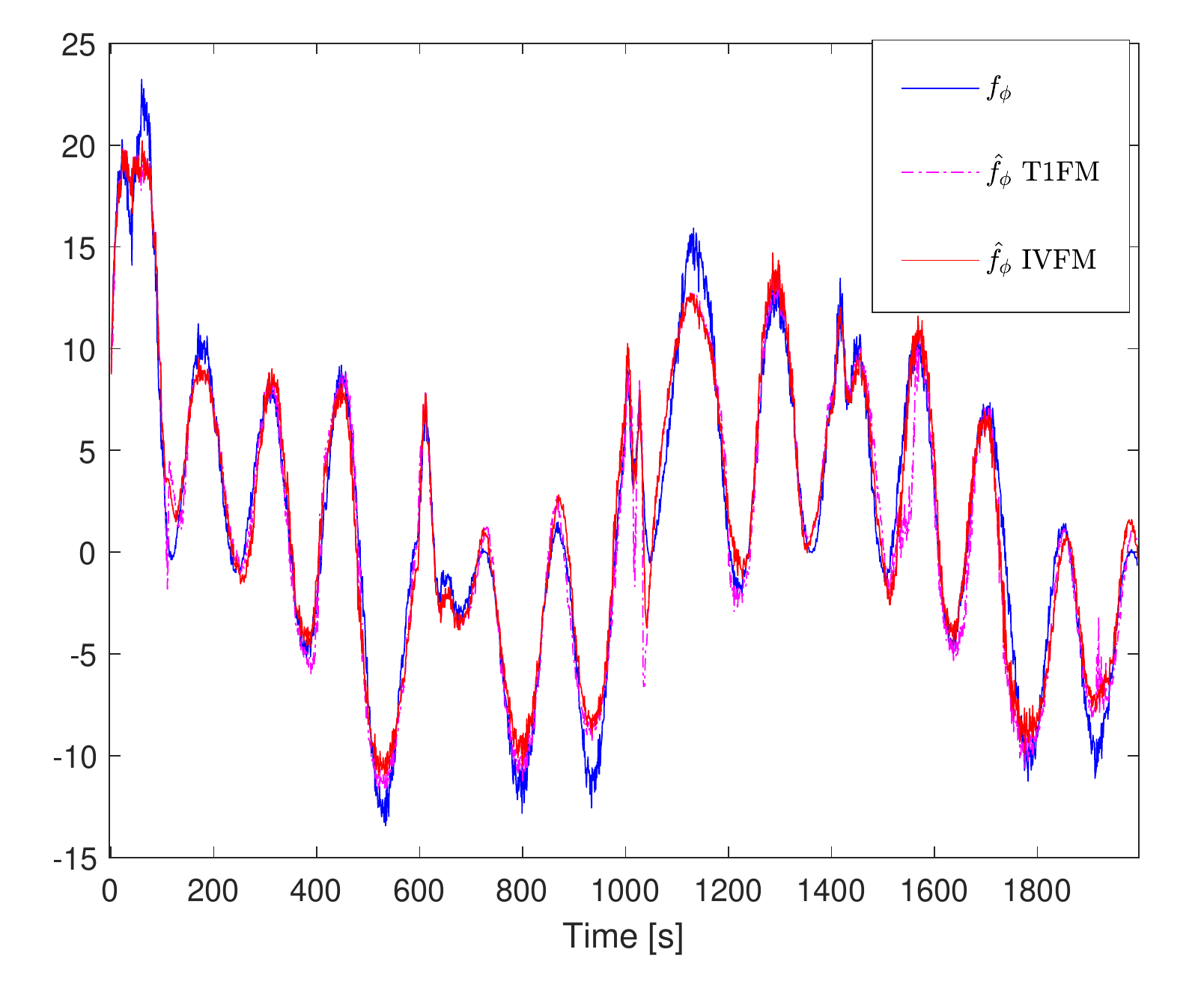}}
\end{center}
\caption{ Comparaison between ${\hat f_{\phi}}({x})$ and ${f_{\phi}}({x})$}\label{fig-eg}
\label{fig7}
\end{figure}

\begin{figure}[!htb]
\begin{center}
%\centerline{\includegraphics[width=4.00in]{figure/fig8.pdf}}
\scalebox{0.6}{\includegraphics{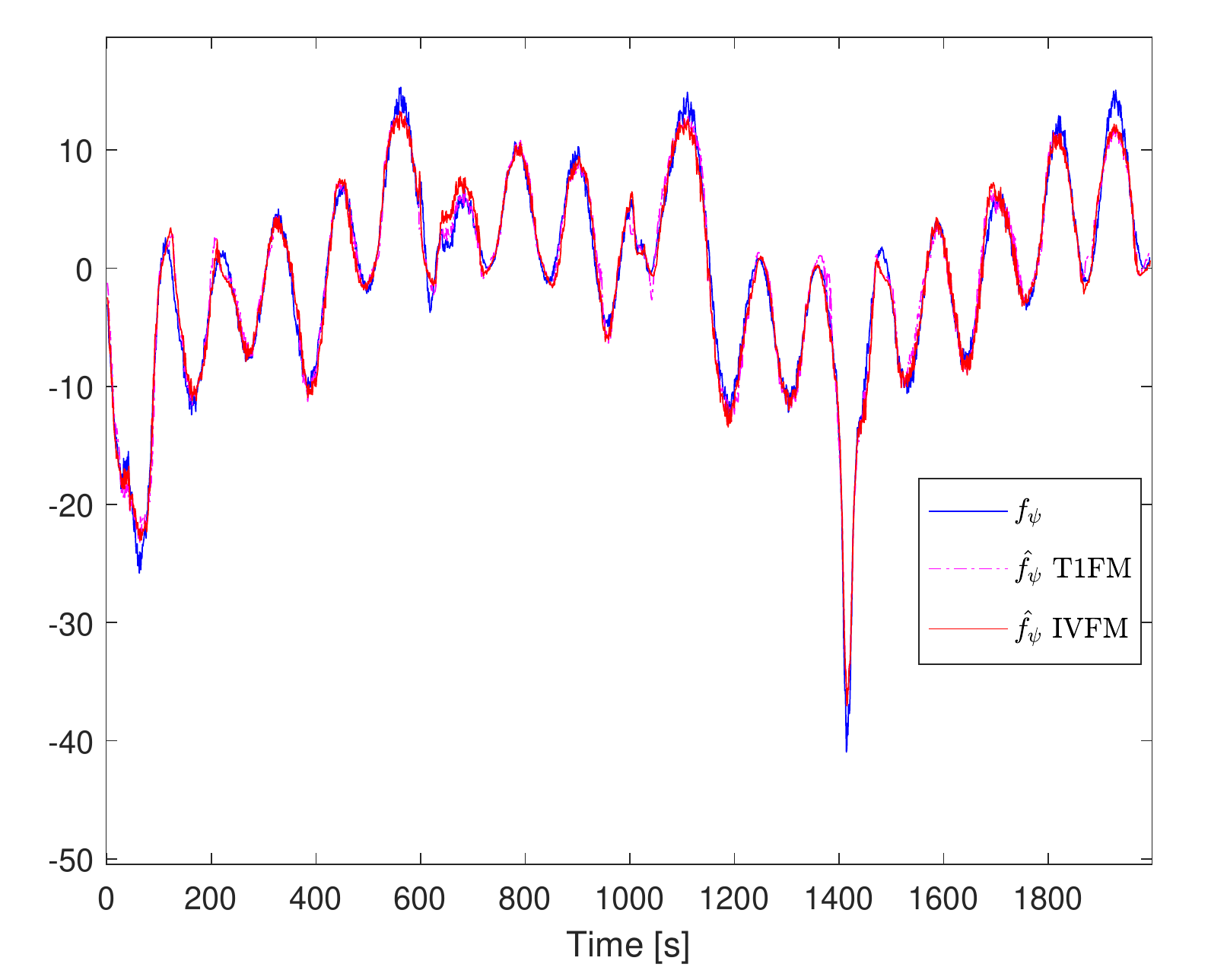}}
\end{center}
\caption{ Comparaison between ${\hat f_{\theta}}({x})$ and ${f_{\theta}}({x})$}
\label{fig8}
\end{figure}

\begin{figure} 
\begin{center}
\scalebox{0.6}{\includegraphics{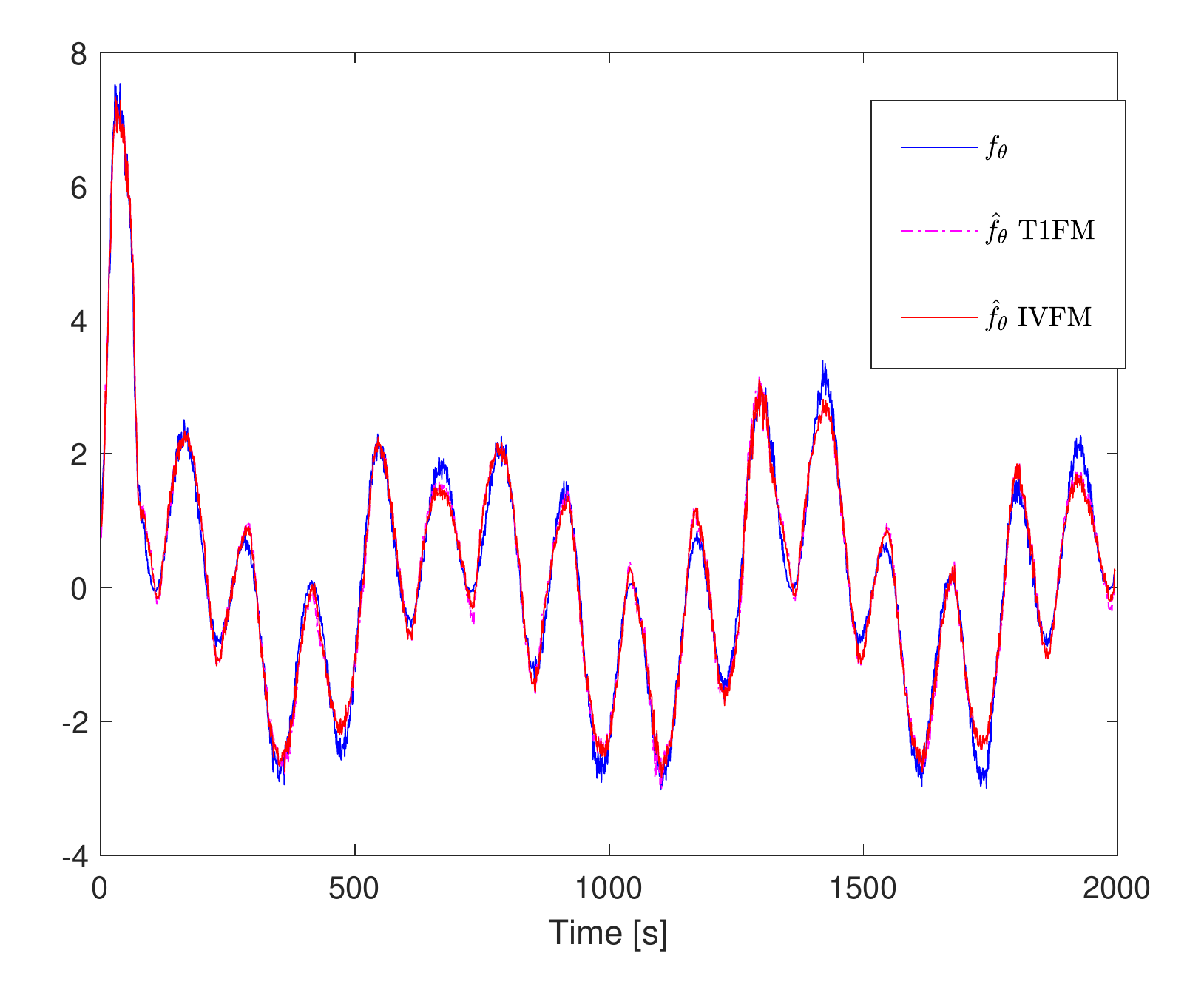}}
%\centerline{\includegraphics[width=4.00in]{figure/fig9.pdf}}
\end{center}
\caption{ Comparaison between ${\hat f_{\psi}}({x}) $ and $ {f_{\psi}}({x})$ }
\label{fig9}
\end{figure}

The validation of fuzzy models obtained (the fuzzy model type-2 ; IFVS and the fuzzy model type-1 ; T1FS) as shown in Fig.~\ref{fig7} ~\ref{fig8} ~\ref{fig9} and the Table~\ref{table_rmse} show that our proposed technique ( IVFM) is more close to real functions of system compared to the type-1 model.

%%%%%%%%%%%% table 2

\begin{table}[t]
\caption{Comparison between RMSE of T1FM and IVFM }
\begin{center}
\label{table_rmse}
\begin{tabular}{c l l l}
%& & \\ % put some space after the caption
\hline
$f_i(x,\Omega)$   & $\hat f_i(x,\theta_{fi})$   & T1FM	& IVFM\\
\hline
${f_{\phi}}(x,\Omega)$ & ${\hat f_{\phi}}({x_{8}},{x_{12}},{\theta _{f_{\phi}}})$ & 1.5958	 & 1.4016 \\
 ${f_{\theta}}(x,\Omega)$&${\hat f_{\theta}}({x_{10}},{x_{12}},{\theta _{f_{\theta}}})$ & 1.1717	 &1.1009 \\
${f_{\psi}}(x,\Omega)$ &${\hat f_{\psi}}({x_{10}},{x_{12}},{\theta _{f_{\psi}}})$ & 0.2072	 &0.1930 \\
\hline
\end{tabular}
\end{center}
\end{table}
%%%%%%%%%%%%%%%%%%%%%%%%%%%%%%% end table 
%$f_{\phi}$, $f_{\theta}$ and $f_{\psi}$ 

%%%%%%%%%%%%%%%% begin figure %%%%%%%%%%%%%%%%%%%

%\begin{figure} 
%\centerline{\includegraphics[width=4.00in]{figure/fig7.pdf}}
%\caption{ Comparaison between ${\hat f_{\phi}}({x})$ and ${f_{\phi}}({x})$}
%\label{fig7}
%\end{figure}

%\begin{figure} 
%\centerline{\includegraphics[width=4.00in]{figure/fig8.pdf}}
%\caption{ Comparaison between ${\hat f_{\theta}}({x})$ and ${f_{\theta}}({x})$}
%\label{fig8}
%\end{figure}

%\begin{figure} 
%\centerline{\includegraphics[width=4.00in]{figure/fig9.pdf}}
%\caption{ Comparaison between ${\hat f_{\psi}}({x}) $ and $ {f_{\psi}}({x})$ }
%\label{fig9}
%\end{figure}
%%%%%%%%%%%%%%%% end figure %%%%%%%%%%%%%%%%%%%

\section{ Indirect Adaptive control}
\label{ind_ctrl}
The initial model of the antecedent part in Eqn.~(\ref{rule}) plays an important role in the structure of the system. The previous clustering algorithm is used to build the initial structure of the unknown part of a nonlinear system with a fuzzy model. The laws of Lyapunov are then applied to adjust the identified model while maintaining system stability. 

Let $\underline{e}(t) = \underline{x} _m(t) - \underline{x} (t)$  the error of tracking, $K = [k_0 , 1]$ a vector of parameters of conception and the equation of sliding is given by: 
\begin{equation}
\underline{e}_{s}(t) = \underline{e}  + {k_0}\dot {\underline{e}} 
\end{equation}
where $ e_s(t)=(e_{s_x},e_{s_y},e_{s_z},e_{s_\theta},e_{s_\varphi},e_{s_\psi}) $ represent the sliding surface.

%%%%%%%%%%%%
\subsection{ Design of fuzzy adaptive SMC control law }
The approximed fonctions $f(x)$ and $g(x)$ with the universelle approximator flous, take the following forme:$\hat f(x,\theta ) = {\theta_f}^T{\varphi_f}(x)$  and $\hat g(x,\theta ) = {\theta _g}^T{\varphi _g}(x)$  With ${\varphi _f}$  and  are the basis of fuzzy vector functions, supposed convenablement fixed by the fuzzy identification presented in ~\ref{IVFM_quad},   and  are the vectors of paramètres ajustables supposed définis respectively in the domaine ${\Omega _f}$and ${\Omega _g}$ . We défine the sub-set ${S_x} \subseteq {\Re ^n}$ as a space in which the trajectoiries of states  $x$ stay inside it in the closed loop. The functions $f(x)$ and $g(x)$ are writen as follow :
   
\begin{equation}
 % \label{eq:t}
  \begin{aligned}
  \hat f(x,\theta _f) &= {f^*}(x,\theta _f^*) + w_f(x) \hfill \\
  \hat g(x,\theta _g) &= {g^*}(x,\theta _g^*) + w_g(x) 
  \end{aligned}
\end{equation}																

with ${w_f}$ and  ${w_g}$ represente errors of appriximation. ${f^ * }(x,\,\theta )$ and ${g^ * }(x,\,\theta )$  are respectivly the optimal parameters of $\hat f(x)$  and $\hat g(x)$ . The values of paramètres ${\theta _f}$  and ${\theta _g}$ minimise respectivly tthe errors of approximation ${w_f}(x)$  and ${w_g}(x)$ . those optimal paramètres satisfy :

\begin{equation}
  %\label{eq:t}
  \begin{aligned}
{\theta_f}^* &= \arg \mathop {\min }\limits_{{\theta _f} \in {\Omega _f}} \left( {{{\sup }_{x \in {S_x}}}\left| {{\theta _f}^T{\varphi _f}(x) - f(x)} \right|} \right) \hfill \\    
{\theta _g}^* &= \arg \mathop {\min }\limits_{{\theta _g} \in {\Omega _g}} \left( {{{\sup }_{x \in {S_x}}}\left| {{\theta _g}^T{\varphi_g}(x) - g(x)} \right|} \right) \hfill \\ 
  \end{aligned}
\end{equation}

We suppose that $\left| {{w_f}(x)} \right| \leq {W_f}$  and $\left| {{w_g}(x)} \right| \leq {W_g}$  where ${W_f}(x)$ and ${W_g}(x)$ are the known upper bound of approximation error. Both fonctions  and  can be choose arbitrairely small. The parametres of errors are :

\begin{equation}
	 {\begin{array}{*{20}{c}}
  {{{\tilde \theta }_f}(t) = {\theta _f}(t) - \theta _f^*} \\ 
  {{{\tilde \theta }_\beta }(t) = {\theta _g}(t) - \theta _g^*} 
\end{array}} 
\end{equation}
Considering the following proposed indirecte fuzzy control 
  
Both terms  ${u_{eq}}$ and ${u_{s}}$ will be calculated with details in the next sections. In the following we note IVFC (Interval value fuzzy control) and IVFM (Interval value fuzzy Model) for the control based on type-2 fuzzy system, in other hand we note T1FC and T1FM for the model based on type-1 fuzzy and Type-2 fuzzy system respectively.

%%%%%%%%%%%%%%%%%%%%%%%%%%%%%%%%%%
\subsubsection{Terme of equivalance control ${u_{eq}}$}
\label{ctrl_equ}

take $\bar e_s(t) = k_{d - 2}e^{\left( {d - 1} \right)}(t) + \cdot\cdot\cdot + {k_0}\dot e(t)$ in order to have ${\bar e_s}(t) = {\dot e_s}(t) - {e^{\left( d \right)}}(t)$. The certainty equivalence control term \cite{[17]}  become as:
\begin{equation}
\label{equiv_1}
              u_{eq} =  - \frac{1}{\hat g(x)}(\hat f(x) + \nu (t))
\end{equation}

where  $\nu (t) = {y_d}^{\left( d \right)} + \gamma {e_s} + {\bar e_s}$   and  $\gamma  > 0$  is a design parameter. The dth derivative of the output error ${e^{\left( d \right)}} = {y_d}^{\left( d \right)} - {y^{\left( d \right)}}$  with $u = {u_{eq}} + {u_s}$, hence :

\begin{equation}
 % \label{eq:t}
  \begin{aligned}
    e^{(d)}& =y_d^{(d)} - f(x) - g(x)u(t)\\
           &= y_d^{(d)} - f(x) - \frac{g(x)}{\hat g(x)}(-\hat f(x) + \nu (t)) -( g(x))u_s
  \end{aligned}
	\label{equation14}
\end{equation}

Note that the two first termes of the Eqn.~(\ref{equation14}):

\begin{equation}
  \begin{aligned}
y_d^{(d)} - f(x) &= y_d^{(d)} - \hat f(x) - f(x) + \hat f(x) \\
                 &= (- \hat f(x) + \nu (t)) - f(x) + \hat f(x) - \gamma e_s - \bar e_s
\end{aligned}
\label{equation15}
\end{equation}

We substitute Eqn.~(\ref{equation15}) in Eqn.~(\ref{equation14}), leads to :

\begin{equation}
 % \label{eq:t}
  \begin{aligned}
e^{(d)} &= \left( {\hat f(x) - f(x)} \right) + \left( {\hat g(x) - g(x)} \right)u_{eq} \\
        & - \gamma {e_s} - {\bar e_s} - g(x){u_s}
  \end{aligned}
\end{equation}

And with ${\bar e_s} = {\dot e_s} - {e^{(d)}}$ , leads to :
\[{\dot e_s} + \gamma {e_s} = \left( {\hat f(x) - f(x)} \right) + \left( {\hat g(x) - g(x)} \right){\text{ }}{u_{eq}} - \left( {g(x)} \right){u_s}\]
%%%%%%%%%%%%%%%
\subsubsection{law of parameters adaptation  }

Consider the lyapunov candidate function:
\begin{equation}
V_{i}=\frac{1}{2}e_s^2 + \frac{1}{2\eta _f}{\tilde \theta _f}^T\tilde \theta _f + \frac{1}{2\eta _g}{\tilde \theta _g}^T\tilde \theta _g
\label{lyap}
\end{equation}
Where ${\eta _f} > 0$ and ${\eta _g} > 0$ are  the parameters of conception. The derivative of Lyapunov function Eqn.~(\ref{lyap}), gives us:

\begin{equation}
\begin{aligned}
\dot V_i &= {e_s}\left(-\gamma {e_s} + (\hat f-f)+ \left(\hat g - g \right)u_{ce} - (g)u_s\right) \\ 
 &+\frac{1}{\eta _f}{\tilde \theta _f}^T{\dot \tilde \theta _f} + \frac{1}{\eta _g}{\tilde \theta _g}^T{\dot \tilde \theta _g} \\
\end{aligned}
\end{equation}

In other hand ,
\begin{equation}
  %\label{eq:t}
\begin{aligned}
\hat f - f &= \theta _f^T\phi _f(x) - {\theta^{*}_{f}}^T\phi _f(x) - {w_f}(x) \\ 
	         &= \tilde \theta _f^T\phi _f(x) - w_f(x) \hfill \\ 
\end{aligned}
\end{equation} 

Same thing with $(\hat g - g)$ hence,

\begin{equation}
  %\label{eq:t}
  \begin{aligned}
{\dot V_i} & = + \left( {{{\tilde \theta }_f}^T{\phi _f} - {w_f} + {{\tilde \theta }_g}^T{\phi _g}{u_{eq}} - w_g{u_{ce}} - g{u_s}} \right){e_s} \\
	     & + \frac{1}{\eta _f}{\tilde \theta _f}^T{\dot \tilde \theta _f} + \frac{1}{\eta _g}{\tilde \theta _g}^T\dot \tilde \theta _g- \gamma {e_s}^2 
  \end{aligned}
\end{equation}

We choose the  following adaptation laws :
\begin{equation}
  %\label{eq:t}
  \begin{aligned}
   \dot \theta _f(t) &=  - {\eta _f}{\varphi _f}(x)e_s \hfill \\
   \dot \theta _g(t) &=  - {\eta _g}{\varphi _g}(x)e_su_{eq}  \hfill \\
  \end{aligned}
\end{equation}

where  ${\varphi _f}(x) = \tfrac{1}{2}\left( {{\xi _l}^f + {\xi _r}^f} \right)$  and ${\varphi _g}(x) = \tfrac{1}{2}\left( {{\xi _l}^g + {\xi _r}^g} \right)$ , the speed of adaptation related to the choice of  the values of ${\eta _f}$ and ${\eta _g}$ . we suppose that the ideal parameters are constants ${\dot \tilde \theta _f} = {\dot \theta _f}$ and ${\dot \tilde \theta _g} = {\dot \theta _g} $. With those conditions we find that:$\frac{1}{{{\eta _g}}}{\tilde \theta _g}^T{\dot \tilde \theta _g} =  - {\tilde \theta _g}^T{\varphi _g}\left( x \right){e_s}{u_{eq}}$  and $\frac{1}{{{\eta _f}}}{\tilde \theta _f}^T{\dot \tilde \theta _f} =  - {\tilde \theta _f}^T{\varphi _f}\left( x \right){e_s}$ hence, 
\begin{equation}
\label{lyap_dot_0}
{\dot V_i} =  - \gamma {e_s}^2+ \left({- {w_f}(x) - {w_g}(x){u_{eq}} - g(x){u_s}} \right){e_s}
\end{equation}

%%%%%%%%%%%%%%%%%%%%%%%%%%%%%%%%%%%%%%%%%%%%%%%%%%%%%%%%%%%%%%%%%%%%%

\subsubsection{Projection Modification to Parameter’s Update Laws  }
The laws of adaptation in the equations (18-19) don’t garanty that ${\theta _f} \in \,{\Omega _f}$  and ${\theta _g} \in \,{\Omega _g}$ . We must use the projection to insure that (e.g. to guaranty that $\hat g\left( x \right) \geq {g_0}$ ). We suppose in particuliar that the i$^{th}$ component of  ${\theta _f}^*$ and ${\theta _f}^*$  is in the known intervalle:  $\theta _{{f_i}}^* \in {\text{[}}{\theta _{{f_i}}}^{\text{min}},{\theta _{f_i}}^{\text{max}}{\text{]}}$   and $\theta _{{g_i}}^* \in {\text{[}}\theta _{{g_i}}^{{\text{min}}},\theta _{{g_i}}^{{\text{max}}}{\text{]}}$  and we want that  and  , we define   as a point in the acceptable region. We can resume this algorithm by using the following rule :
 $ \text{if }{\theta _f}(t) \notin \,[{\theta _f}^{\min },{\theta _f}^{\max }] $ and $ \theta _f^{ud}({\theta _f} - \theta _f^m) \geq 0 $ then $ {\dot \theta _f}(t) = 0 $ else  $ \dot \theta _f(t) =  - {\eta _f}{\varphi _f}e_{\text{s}}u_{\text{eq}} $ \\
In the stability analysis, this projection modification to the update laws will always result in a parameter estimation error that will decrease ${ V_i}$ at least as much as if the projection were not used; hence, the right-hand side of Eqn.~(\ref{lyap_dot_0}) will overbound the  ${\dot V_i}$ that would result if projection is used. For this reason, we conclude that:
\begin{equation}
{\dot V_i} \leq  - \gamma {e_s}^2\, + \left( { - {w_f}(x) - {w_g}(x){u_{ce}} - g(x){u_s}} \right){e_s}
\label{dvi}
\end{equation}
%%%%%%%%%%%%%%%%%%%%%%%%%%%%%%%%%%%%%%%%%%%%%%%%%%%%%%%%%%%%%%%%%%%%%

\subsubsection{Sliding Mode Control Term }
\label{ctrl_s}

To ensure that Equation Eqn.~(\ref{dvi}) is less than or equal to zero, we choose:

$$ {u_s} = \eta (x)\operatorname{sgn} ({e_s}) = \frac{1}{g_0}\left( {{W_f}(x) + {W_g}(x)\left| u_{eq} \right|} \right)\operatorname{sgn} (e_s) $$
We see that  : $ -(w_f(x) + w_g(x)u_{ce}){e_s} \leq ( {| w_f(x)| + |w_g(x)u_{ce}|})|e_s| $ with $ \frac{g(x)}{g_0} \geq 1 $ hence:
\begin{equation}
\begin{aligned}
\dot V_i & \leq  - \gamma e_s^2 + (|{w_f}(x)| + |{w_g}(x)u_{eq}|)|e_s|  \\
              & - {e_s}\frac{g}{g_0}(W_f(x) + W_g(x)|u_{eq}|)\operatorname{sgn} (e_s)  \\
\end{aligned}
\end{equation}

\begin{equation}
\begin{aligned}
{\dot V_i} & \leq  - \gamma {e_s}^2 + \left| {{w_f}\left( x \right)\left| {\left| {e_s} \right|}+ \right|{w_g}\left( x \right){u_{eq}}} \right|\left| {e_s} \right| \\
 &- {e_s}sgn\left( {e_s} \right){W_f}\left( x \right) - {e_s}sgn\left({e_s}\right){W_g}\left( x \right)\left| u_{eq} \right|
\end{aligned}
\end{equation}
where $ |e_s| = e_ssgn(e_s)$ ,$\forall {e_s} \ne 0:$  and $|w_g(x)| \leq W_g(x) $  then :
 $$ |{w_f}(x)||e_s| - {e_s}sgn({e_s}){W_f}(x) = |e_s|(|w_f(x)| - W_f(x)) \leq 0 $$
%\begin{equation}
%\label{eq:t}
$$
\begin{aligned}
\left| {{w_g}(x){u_{eq}}} \right|\left| {{e_s}} \right| &- {e_s}sgn({e_s}){W_g}(x)\left| {{u_{eq}}} \right| =\\
& \left| {{e_s}} \right|(\left| {{w_g}(x){u_{eq}}} \right| - {W_g}(x)\left| {{u_{eq}}} \right|) \leq 0
\end{aligned}
$$
%\end{equation} 																									
\begin{equation}
\dot V_i \leq  - \gamma {e_s}^2 
\end{equation}
because $\gamma {e_s}^2 \geq 0 $ , this show that ${\dot V_i}$  which is a measure of the tracking error and parameter estimation error, is a nonincreasing function of time. Notice that $\gamma > 0$   has an influence on how fast $V_i \mapsto 0$. By picking $\gamma$ larger you will often get faster convergence of the tracking error. 

%%%%%%%%%%%%%%%%%%%%%%%%%%%%%%%%%%%%%%%%%%%%%%%%%%%%%%%%%%%%%%%%%%%%%%%%
\subsubsection{Asymptotic Convergence of the Tracking Error and Boundedness of Signals
 }

Due the hypothysis considered (that the reference signals are bounded, $x$ is measurable, $d = n$, and the projection ensures that the ${u_{eq}}$ term is well-defined), the following statements hold:
\begin{enumerate}
	\item 	The system output $y,\dot y,...,{y^{(d - 1)}}$ are bounded 
	\item The control signal $u$, ${u_{eq}}$ et ${u_{s}}$ are bounded. 
	\item The parameteres ${\theta _g}(t)$ and ${\theta _g}(t)$ are bounded. 
	\item  The magnitude of the output error decreases at least asymptotically to zero  ( $\mathop {\lim }\limits_{t \to \infty } e(t) = 0$) 
\end{enumerate}
 
To prove this, Note that since $V_i$  is positive and ${\dot V_i} \leq  - \gamma {e_s}^2$ . We know that ${e_s},{\theta _f}$ and ${\theta _g}$ are bounded. since $e_s$ is bounded and $y_r$ and its derevatives are bounded. We know that $y,\dot y,...,{y^{(d - 1)}}$ are bounded.  Hence, $x$ are bounded. Hence,  $f(x),\hat f(x),g(x)$  and  $\hat g(x)$ are bounded. Since $x$ is bounded  and $\hat g(x) \geq {g_0}$ ,  $u_{eq}$ and $u_s$  and hence $u$, are bounded. Next, note that :
 
 \begin{equation}
	 \int\limits_0^\infty  {\gamma e_s^2dt}  \leq - {\int V _i}dt = {V_i}(0) - {V_i}(\infty )
 \end{equation}
																	
This establishes that  ${e_s} \in {L_2}{\text{ ou }}{L_2} = \left\{ {z(t):\int_0^\infty  {{z^2}(t)} dt < \infty } \right\}$ since ${V_i}\left( 0  \right)$ and ${V_i}\left( \infty  \right)$  are bounded. Note that via the equation (15), ${\dot e_s}$  is bounded. Hence, because  ${\dot e_s}$ and  ${e_s}$ are bounded and ${e_s} \in {L_2}$ ,  and via the barbarat’s lemma we have $\mathop {\lim }\limits_{t \to \infty } {e_s}(t) = 0$ .  Hence $\mathop {\lim }\limits_{t \to \infty } e(t) = 0$. \\
It is possible to reduce the chattering that can be result in the sliding mode term of control, we introduce  ${{\mathbf{e}}_\varepsilon } = (e_{\varepsilon _x},e_{\varepsilon _y},e_{\varepsilon _z},e_{\varepsilon _\theta },e_{\varepsilon _\varphi},e_{\varepsilon_\psi })$  we can use the smooth function $ \textnormal{sgn}({e_s})$ \cite{[18]}. In this case, however, you only get convergence to an neighborhood of ${e_s} = 0$ . To prove the stability, we define the error ${e_\varepsilon }$.

 \begin{equation}
 {e_\varepsilon } = {e_s} - \varepsilon {sat}(\frac{e_s}{\varepsilon })
\end{equation}

 %\begin{equation}
 where
$$
 {sat(x)} = \left\{ \begin{array}{*{20}{c}}
 1{\textnormal{   si}}\,\,1 \leq x \hfill \\
 x{\textnormal{   si}}\,\, - 1 < x < 1 \hfill \\
 - 1{\textnormal{ si}}\,\,x \leq - 1 \hfill 
\end{array}
\right.  
$$
%\end{equation}

where ${e_\varepsilon }$  represent the distance between ${e_s }$ and the desired “boundary layer,”, so the ${e_\varepsilon}=0$ when ${e_s }$ is inside the layer. The equivalent control is defined to be:
\[{u_{eq}} = - \tfrac{1}{\hat g(x)}(\hat f(x) + {\nu _\varepsilon }(t))\,\,{\text{with}}\,\,{\nu _\varepsilon }(t) = {y_d}^{(n)} + \gamma {e_\varepsilon } + {\bar e_s}\]
we consider the following lyapunouv function:
 \[{V_{i}} = \frac{1}{2}{e_\varepsilon }^2 + \frac{1}{{2{\eta _f}}}{\tilde \theta _f}^T{\tilde \theta _f} + \frac{1}{{2{\eta _g}}}{\tilde \theta _g}^T{\tilde \theta _g}\]
And the Adaptation laws became :
%$$
\begin{equation}
\begin{aligned}
{\dot \theta }_f(t) &=  - {\eta _f}{\varphi _f}(x){e_\varepsilon }\\
{\dot \theta }_g(t) &=  - {\eta _g}{\varphi _g}(x){e_\varepsilon }{u_{eq}}
\end{aligned}
\label{adapt_law_2}
\end{equation}
%$$
with those laws the derived lyapunov functions:
 \[{\dot V_i} \leq  - \gamma {e_\varepsilon }^2\, + \left( {\left| {{w_f}(x)} \right| + \left| {{w_g}(x){u_{ce}}} \right|} \right)\left| {{e_\varepsilon }} \right| - g(x){u_s}{e_\varepsilon }\]
the terme $u_s$ in Eqn.~(\ref{sliding_2}) become a control with smooth action.
 
\begin{equation}
\label{sliding_2}
{u_s} = \frac{1}{{{g_0}}}\left( {{W_f}(x) + {W_g}(x)\left| {{u_{eq}}} \right|} \right){\text{sat}}(\frac{{{e_s}}}{\varepsilon })\end{equation}                     
with $ {e_\varepsilon }{\text{sat}}(\frac{e_s}{\varepsilon }) = \left| e_\varepsilon  \right| $, we can proof that $\dot V_i \leq - \gamma {e_\varepsilon }^2 $ , that’s make $\left| {{e_\varepsilon }} \right|$  asymptotically stable  , in addition ${e_s}$ will converge to the  $\varepsilon$-neighbor of $ e_s =0 $, and $e$  will converge to 0.

%%%%%%%%%%%%%%%%%%%%%%%%%%%%%
\subsection{Control of position and attitude of Quadrotor}
\label{sect_ctrl_quad}

%%%%%%%%%%%%%%%%%%%%%%%%%%%%%11111111111111111%%%%%%%%%%%%%%%%%%%%%%%%%%%%%%%%%
{\bfseries Proposition 1}: for the conception of attitude controller described by the following sub-systems ~(\ref{sub1}),~(\ref{sub2}) and ~(\ref{sub3}).

we propose a sliding mode controller for every subsystem $i$ where $i = \left\{\phi ,\theta, \psi \right\}\,$ as follow:
\begin{equation}
{U_i} = {u_{eq_{i}}} + {u_{s_i}}
\label{ctrl1}
\end{equation}

 where :
 $$
 \begin{aligned}
{u_{e{q_i}}} &=  - \frac{1}{{{{\hat g}_i}(x)}}\left( {{{\hat f}_i}(x,{\theta _{{f_i}}}) + \ddot y_{{d_i}}+ \gamma {e_{{\varepsilon _i}}} + {{\bar e}_{{s_i}}}} \right) \\
{u_{{s_i}}} &= \frac{1}{g_{0i}}\left( {{W_{{f_i}}}(x) + {W_{{g_i}}}(x)|u_{e{q_i}}|} \right)sat(\frac{e_{s_i}}{\varepsilon })
 \end{aligned}
$$
where ${\bar e_s} = {k_i}e$ , ${y_{{d_i}}} = \left\{ {{\phi _d},{\varphi _d},{\theta _d}} \right\}\,$, $d = 2$ ,$W_{f_i}(x)$ and  $W_{g_i}(x)$ are the errors bounds of the approximation of  $f_i$ and $g_i$ respectively. with  $g_{0_i}$  is the minimum value of $b_i$ . the control input  Eqn.~(\ref{ctrl1}) ensures the asymptotic convergence of $\phi ,\theta $  and $\psi$ to ${\phi _d},{\theta _d},\,{\psi _d}$ respectively. \\

%%%%%%%%%%%%%%%%%%%%%%%%proof%%%%%%%%%%%%%
{\bfseries Proof }: similarly, to sections ~\ref{ctrl_equ} and ~\ref{ctrl_s} , we take $n = 2$, the equations Eqn.~(\ref{equiv_1}) and Eqn.~(\ref{sliding_2}) became:

\begin{multline}
 {U_{i}} = \frac{1}{{\hat b}_{i}}\left( - {{\hat f}_{i}}\left( x,\varphi _{f_{i}} \right) + \gamma {e_{\varepsilon i}} +  {{\bar e}_{si }} + \ddot y_{i_d} \right) \\
	      	 + {\frac{1}{g_{0{i}}}(W_{f_{\phi}}(x) + W_{g_{i}}(x)|u_{e{q_{i}}}|) }{\text{sat}}(\frac{e_{s\phi }}{\varepsilon })
\end{multline}

using the equations in Eqn.~(\ref{adapt_law_2}), the adaptations laws can be written as follow:
        
%\[\begin{array}{*{20}{l}}
  %\begin{gathered}
  %{{\dot \theta }_{f1}} =  - {\eta _{f_1}}{\varphi _{f_1}}\left( {\underline{x} } \right){e_{\varepsilon \phi }} \\ 
  %{{\dot \theta }_{f2}} =  - {\eta _{f_2}}{\varphi _{f_2}}\left( {\underline{x} } \right){e_{\varepsilon \theta }} \hfill \\
  %{{\dot \theta }_{f3}} =  - {\eta _{f_3}}{\varphi _{f_3}}\left( {\underline{x} } \right){e_{\varepsilon \psi }}\,\, \hfill \\ 
%\end{gathered}  
%\end{array}\]

\begin{equation}
\begin{aligned}
{\dot \theta }_{f{i}}&=  - {\eta _{f_{i}}}{\varphi _{f_{i}}}\left( {{x} } \right){e_{\varepsilon_i }} \hfill \\
{\dot \theta }_{g{i}} &= - {\eta _{g_{i}}}{e_{\varepsilon i }} u_{eq_{i}}\hfill \\
\end{aligned}
\label{adapt_1}
\end{equation}

Notice that the functions $g_i$ in Eqn.~(\ref{adapt_1}) is considered as a constants that's why it is approximated with a constant and not a fuzzy function, so we take ${\varphi _{f_i}} = 1 $ .

%%%%
%
%\begin{equation}
%\left. \begin{array}{*{20}{c}}
%{\dot \theta }_{f{\phi}} =  - {\eta _{f_{\phi}}}{\varphi _{f_{\phi}}}\left( {{x} } \right){e_{\varepsilon \phi }} \\ 
%{\dot \theta }_{f{\theta}}= - {\eta _{f_{\theta}}}{\varphi _{f_{\theta}}}\left( {{x} } \right){e_{\varepsilon \theta }} \hfill \\
%{\dot \theta }_{f{\psi}} =  - {\eta _{f_{\psi}}}{\varphi _{f_{\psi}}}\left( {{x} } \right){e_{\varepsilon \psi }}\,\, \hfill 
%\end{array}  
%\right|\begin{array}{*{20}{c}}
  %{{\dot \theta }_{g{\phi}}} =  - {\eta _{g_{\phi}}}{e_{\varepsilon \phi }} u_{eq_{{\phi}}}\\ 
  %{{\dot \theta }_{g{\theta}}}= - {\eta _{g_{\theta}}}{e_{\varepsilon \theta }}u_{eq_{{\theta}}} \hfill \\
  %{{\dot \theta }_{g{\psi}}} =  - {\eta _{g_{\psi}}}{e_{\varepsilon \psi }}u_{eq_{{\psi}}}\,\, \hfill 
%\end{array}
%\label{adapt_1}
%\end{equation}
%Notice that the functions $g_i$ in Eqn.~(\ref{adapt_1}) is considered as a constants that's why it is approximated with a constant and not a fuzzy function, so we take ${\varphi _{f_i}} = 1 $ .

%%%%%%%%%%%%%%%%%%%%%%%%%%%%%%%222222222222222222%%%%%%%%%%%%%%%%%%%%%%%%%%%%%%%%%%%%%%%%%%%%%%%%%%%%%%%

{\bfseries Proposition 2} : for the altitude controller, the subsystem that governs the movement $z(t)$ is :
 \[\left( {\begin{array}{*{20}{c}}
  {{{\dot x}_7}} \\ 
  {{{\dot x}_8}} 
\end{array}} \right) = \left( {\begin{array}{*{20}{c}}
  {\dot z} \\ 
  {{f_z} - U{g_z}{U_z}} 
\end{array}} \right)\,\]
where $U = (\cos \phi \cos \theta )$ , ${g_z} = \frac{1}{m}$ and ${f_z}$ is the gravity $g$.
The sliding mode control Eqn.~(\ref{ctrl2}) ensure the convergence of $z(t)$ toward the consign $z_d $

%\begin{equation}
 %% \label{eq:t}
  %\begin{aligned}
%{U_z} &= \frac{1}{{U{g_z}}}( - {\hat f_z}(x) + {\ddot z_d} + \gamma {e_{\varepsilon z}} + {\bar e_{sz}}) \\
%& + \frac{1}{g_0} \left( {{W_{f_z}}(x)/(U) + {W_{g_z}}(x)\left| {u_{eq_z}} \right|} \right){\text{sat}}(\frac{e_{sz}}{\varepsilon})
  %\end{aligned}
	%\label{ctrl2}
%\end{equation}

\begin{multline}
{U_z} = \frac{1}{{U{g_z}}}( - {\hat f_z}(x) + {\ddot z_d} + \gamma {e_{\varepsilon z}} + {\bar e_{sz}}) \\
 + \frac{1}{g_0} \left( {{W_{f_z}}(x)/(U) + {W_{g_z}}(x)\left| {u_{eq_z}} \right|} \right){\text{sat}}(\frac{e_{sz}}{\varepsilon})
\label{ctrl2}
\end{multline}

Where ${W_{f_z}}(x)$  and ${W_{g_z}}(x)$  are the bounded limit of error of the approximation of ${f_z}$ and ${g_z}$  respectively\\
%& 
%%%%%%%%%%%%%%%%%%%%proof%%%%%%%~\ref{ctrl_equ} and ~\ref{ctrl_s} %%%%%%%%%%%%%%%%%%%%%%%%
{\bfseries Proof} : following the same procedure in the sections ~\ref{ctrl_equ} and ~\ref{ctrl_s}, considering  ${\ddot z} = {f_4}\left( x \right) - Ug_z{U_z}(t)$ where  , the term of equivalent control  is defined as :

\[{u_{e{q_z}}} =  - \tfrac{1}{U{g_z}}\left( {{{\hat f}_z}(x|{\theta _{f_z}}) + {{\ddot z}_d} + \gamma {e_{\varepsilon _z}} + {{\bar e}_{s_z}}} \right)\]

where $\nu (t) = {\ddot z_d} + \gamma e_{s_z} + {\bar e_{s_z}}$ and $\gamma  > 0$

The 2$^{end}$ derivative of the output error, along with the equation Eqn.~(\ref{equation14})  we can write:

\begin{equation}
\label{ddot_e}
\ddot e = {\ddot z_d} - {f_z}(x) - \tfrac{g_z}{{\hat g}_z}\left(- {{\hat f}_z}(x) + \nu (t) \right) - U{g_z}{u_s}
\end{equation}

Take the two first terms as follow:
\[{\ddot z_d} - {f_z}(x) = \left( - {{\hat f}_z}(x) + \nu (t) \right) - {f_z}(x) + {\hat f_z}(x) - \gamma {e_{{s_z}}} - {\bar e_{{s_z}}}\]

we substitute both terms in the equation Eqn.~(\ref{ddot_e}), using ${\bar e_s}= {\dot e_s}-{e^{(d)}}$  we find:

\[{\dot e_s} + \gamma {e_s} = (\hat f_z - f_z) + (\hat g_z - g_z)U{u_{eq_z}} - Ug{u_s}\]
The candidate function of Lyapunov ${V_i}$ :
\[{V_{z}} = \frac{1}{2}{e_{{\varepsilon _z}}}^2 + \frac{1}{{2{\eta _f}}}{\tilde \theta _{{f_z}}}^T{\tilde \theta _{{f_z}}} + \frac{1}{{2{\eta _g}}}{\tilde \theta _{{g_z}}}^T{\tilde \theta _{{g_z}}}\]
where ${\eta _f} > 0$ and ${\eta _f} > 0$ are the adaptation gain. the derivative of $V_i$ gives us:     

\begin{equation}
\label{lyap2_dot}
\begin{aligned}
\dot V_z &= ({{{\tilde \theta }_{f_z}}^T{\phi _f} - {w_{f_z}} + U{{\tilde \theta }_{g_z}}^T{\phi _g}u_{eq_z}- U{w_g}{u_{eq_z}} - Ug{u_s}})e_{\varepsilon _z} \\
   & - \gamma e_{{\varepsilon _z}}^2 + \frac{1}{{{\eta _f}}}{\tilde \theta _f}^T{\dot \tilde \theta _f} + \frac{1}{{{\eta _g}}}{\tilde \theta _g}^T{\dot \tilde \theta _g}
\end{aligned}
\end{equation}

Suppose that those ideal parameters are constants ${\dot{\tilde \theta} _f} = {\dot \theta _f}$ and ${\dot {\tilde {\theta}} _g} = {\dot \theta _g}$, and the chosen adaptation law is:

\[\begin{array}{*{20}{l}}
  {{\dot \theta }_{f_z}}\left( t \right) =  - {\eta _{f_z}}{\varphi _{{f_z}}}\left( x \right){e_{{\varepsilon _z}}}\\ 
  {{\dot \theta }_{g_z}}\left( t \right) =  - {\eta _{g_z}}{\varphi _{{g_z}}}\left( x \right){e_{{\varepsilon _z}}}{u_{e{q_z}}}{\text{(}}\cos \phi \cos \theta )
\end{array}\]

That’s lead to: $ \frac{1}{\eta _g}{\tilde \theta _g}^T{\dot \tilde \theta _g} =  - {\tilde \theta _g}^T{\varphi _g}\left( x \right){e_{{\varepsilon _z}}}{u_{eq}}$ and $\tfrac{1}{{{\eta _f}}}{\tilde \theta _f}^T{\dot \tilde \theta _f} =  - {\tilde \theta _f}^T{\varphi _f}\left( x \right){e_{\varepsilon _z}}$ , so the equation Eqn.~(\ref{lyap2_dot}) can be written as:

\[{\dot V_z} =  - \gamma {e_{\varepsilon _z}}^2+ \left( { - {w_f}(x) - U{w_g}(x){u_{eq_z}} - U{g(x)}{u_s}} \right){e_{{\varepsilon _z}}}\,\]

To analyze the stability, we conclude that:

\begin{equation}
\label{lyap_dot2}
{\dot V_z} \leq  - \gamma {e_{{\varepsilon _z}}}^2+ \left( { - {w_{f_z}}(x) - U{w_{g_z}}(x){u_{eq_z}} - U{g(x)}{u_s}} \right){e_{{\varepsilon _z}}}
\end{equation}

To ensure that the equation Eqn.~(\ref{lyap_dot2})  must be less than or equal to zero, we choose:

\[{u_s} = \tfrac{1}{g_0}\left( \tfrac{{W_{f_z}}(x)}{U_1} + {W_{g_z}}(x)\left| {{u_{eq_z}}} \right| \right){\text{sat}}(\tfrac{{{e_s}}}{\varepsilon }) = \eta (x){\text{sat}}(\tfrac{e_s}{\varepsilon })\]

We note that: $- \left( {{w_f}(x) + {w_g}(x){u_{eq}}} \right){e_s} \leq \left( {\left| {{w_f}(x)} \right| + \left| {{w_g}(x){u_{eq}}} \right|} \right)\left| {{e_s}} \right|$ , so: 

\begin{multline}
\dot V_z =  - \gamma {e_s}^2 + (- {w_{f_z}}(x) - U{w_{g_z}}(x){u_{eq_z}} \\
 {-U \tfrac{g_z}{g_{0_z}}(\tfrac{W_{f_z}(x)}{U} + {W_{g_z}}(x) |u_{eq_z}|})\text{sat}(\tfrac{e_s}{\varepsilon}))e_\varepsilon 
\end{multline}

%\begin{equation}
%%\label{eq:t}
%\begin{aligned}
%\dot V_z &=  - \gamma {e_s}^2 + \left(- {w_{f_z}}(x) - U{w_{g_z}}(x){u_{eq_z}} \\
         %&{- U \tfrac{g_z}{g_0_z}(\tfrac{{W_{f_z}}(x)}{U} + {W_{g_z}}(x) |{u_{eq_z}}|}){\text{sat}}(\tfrac{e_s}{\varepsilon })} \right){e_\varepsilon }
%\end{aligned}
%\end{equation}

\begin{equation}
\label{inequa_lyap3_dot}
\begin{aligned}
{\dot V_z} &\leq  - \gamma e_{{\varepsilon _z}}^2 + \left( {\left| {{w_{f_z}}(x)} \right| + \left| {U{w_{g_z}}(x){u_{eq_z}}} \right|} \right)\left| {e_{\varepsilon _z}} \right| \\
          &- U{e_{{\varepsilon _z}}}g_z(x)( {\tfrac{1}{g_{0_z}}\left({\tfrac{{W_{f_z}}(x)}{U}+ {W_g}(x)\left| {{u_{eq_z}}} \right|} \right){\text{sat}}(\tfrac{e_s}{\varepsilon })} )
\end{aligned}
\end{equation}

Now, we consider the last term of the equation Eqn.~(\ref{inequa_lyap3_dot}) and because  $\tfrac{{g_z(x)}}{g_{0_z}} \geq 1$  so :

\begin{equation}
%\label{eq:t}
\begin{aligned}
\dot {V_i} & = -\gamma {e_{\varepsilon _z}}^2 + |{{w_{f_z}}(x)| |{e_s}|- {e_\varepsilon }sat(\tfrac{e_\varepsilon}{\varepsilon}){W_{f_z}}(x)} \\
           &  + |U{w_{g_z}}( x )u_{eq_z}| |{e_\varepsilon }| - U{e_\varepsilon }sat(\tfrac{e_\varepsilon}{\varepsilon}){W_{g_z}}(x)|{u_{eq_z}}|
\end{aligned}
\end{equation}

with $|{e_\varepsilon }| = {e_s}sat({e_s}/\varepsilon)$ , $|{w_f}(x)| \leq {W_f}(x)$ and  $|{w_g}(x)| \leq {W_g}(x)$ so:

\begin{equation}
\begin{aligned}
|{w_{f_z}}(x)||{e_\varepsilon }| &- {e_\varepsilon }sat(\frac{e_s}{\varepsilon }){W_f}\left( x \right) \\ 
&= \left| {{e_\varepsilon }} \right|\left( {\left| {{w_{f_z}}\left( x \right)} \right| - {W_{f_z}}\left( x \right)} \right) \leq 0 \\
\end{aligned}
\end{equation}

%\begin{equation} \label{eq1}
%\begin{split}
%A & = \frac{\pi r^2}{2} \\
 %& = \frac{1}{2} \pi r^2
%\end{split}
%\end{equation}

 \begin{multline}
|U{w_{g_z}}(x){u_{eq_z}}||{e_\varepsilon }| - U{e_\varepsilon }{\text{sat}}(\tfrac{e_s}{\varepsilon } ){W_{g_z}}(x)|u_{eq_z}|=\\ \left\{
 \begin{array}{l l}
&U \left| {e_\varepsilon } \right|\,\left( {\left| {{w_{g_z}}{u_{eq_z}}} \right| - {W_{g_z}}\left| {u_{eq_z}} \right|} \right) \leq 0 {\text{ if}}\,\,U \geq 0 \\
&\left| U \right|\left| {{e_\varepsilon }} \right|\,\left( {\left| {{w_{g_z}}{u_{eq_z}}} \right| + {W_{g_z}}\left| {u_{eq_z}} \right|} \right) \leq 0 {\text{ if}}\,\,U < 0 \\
 \end{array} \right.
 \end{multline}

%\begin{equation}
%\begin{aligned}
%|U{w_{g_z}}(x){u_{eq_z}}||{e_\varepsilon }| - U{e_\varepsilon }{\text{sat}}(\tfrac{e_s}{\varepsilon } ){W_{g_z}}(x)|u_{eq_z}| =& \hfill \\  
%\left\{ {\begin{array}{*{20}{c}}
  %{&=        U  \left|       {e_\varepsilon } \right|\,\left( {\left| {{w_{g_z}}{u_{eq_z}}} \right| - {W_{g_z}}\left| {u_{eq_z}} \right|} \right) \leq 0 {\text{ if}}\,\,U \geq 0} \\ % \Rightarrow U = \,\,\,\left| U \right|}
  %{&= \left| U \right|\left| {{e_\varepsilon }} \right|\,\left( {\left| {{w_{g_z}}{u_{eq_z}}} \right| + {W_{g_z}}\left| {u_{eq_z}} \right|} \right) \leq 0 {\text{ if}}\,\,U < 0  } %\Rightarrow U =  - \left| U\right|
%\end{array}} \right.
%\end{aligned}
%\end{equation}

\begin{multline}
{\dot V_z} =  - \gamma {e_\varepsilon }^2 + \left| {e_\varepsilon } \right|\left( {\left| {{w_{f_z}}\left( x \right)} \right| - {W_{f_z}}\left( x \right)} \right)\\
          + \left| {u_{eq_z}} \right|\left| {e_\varepsilon } \right|\left( {\left| {U{w_{g_z}}(x)} \right| - U{W_{g_z}}(x)} \right)
\end{multline}

Since $\gamma {e_\varepsilon }^2 \geq 0$ this shows that ${V_z}$ is a non-increasing function of time.

%%%%%%%%%%%%%%%%%%%%%%%%333333333333333%%%%%%%%%%%%%%%%%%%%%%%%%%%%%%%%%%%%%
{\bfseries Proposition 3} : from the Eqn.~(\ref{sys}), the subsystems that govern the dynamics of position $(x , y)$ is given by:

\begin{equation}
\left( {\begin{array}{*{20}{c}}
  {{\dot x}_9} \\ 
  {{\dot x}_{10}} 
\end{array}} \right) = \left( {\begin{array}{*{20}{c}}
  {\dot x} \\ 
  {{u_x}\tfrac{1}{m}{U_z}} 
\end{array}} \right)
\end{equation}
\begin{equation}
\left( {\begin{array}{*{20}{c}}
  {\dot x}_{11} \\ 
  {\dot x}_{12} 
\end{array}} \right) = \left( {\begin{array}{*{20}{c}}
  {\dot y} \\ 
  {{u_y}\tfrac{1}{m}{U_z}} 
\end{array}} \right)
\end{equation}

The following proposed indirect fuzzy adaptative control $u_x$ and $u_y$ Eqn.~(\ref{ctrl3}): 
\begin{equation}
%\label{eq:t}
\begin{aligned}
  {u_x} &= \frac{\hat m}{U_z}({x_d}^{\left( n \right)} + \gamma {e_{\varepsilon _x}} + {{\bar e}_{sx}}) + {\eta _x}(x){\text{sat}}(\tfrac{e_{s_x}}{\varepsilon }) \hfill \\
  \,{u_y} &= \frac{\hat m}{U_z}({y_d}^{\left( n \right)} + \gamma {e_{\varepsilon _y}} + {{\bar e}_{sy}}) + {\eta _y}(x){\text{sat}}(\tfrac{e_{s_y}}{\varepsilon }) \hfill \\ 
\end{aligned}
\label{ctrl3}
\end{equation}

Ensure the stability toward the desired trajectory asymptotically.

 Where  ${\eta _x}(x) = \tfrac{1}{g_{05}}\left( { + {W_{g_5}}(x)\left| {{u_{eq}}} \right|} \right)$   and  ${\eta _y}(x) = \tfrac{1}{{{g_{06}}}}\left( { + {W_{g_6}}(x)\left| {{u_{eq}}} \right|} \right)$ with  ${W_{g_5}}(x)$  and ${W_{g_6}}(x)$  are the known bounds of approximation errors of  functions: ${g_5}$  and ${g_6}$  respectively.

%%%%%%%%%%%%%%%%%%%prooof %%%%%%\right\}\,

{\bfseries proof }: for the subsystem that govern the variable $x$, we have : $\hat g = \frac{1}{m} \to {u_x}\frac{1}{m}{U_1} = {u_x}{\hat g_x}{U_z}$   and by the choice of the candidate function of Lyapunov ${V_i}$ where $i= \left\{x , y \right\}\, $ :
\[{V_{i}} = \frac{1}{2}{e_{{\varepsilon _i}}}^2 + \frac{1}{{2{\eta _f}}}{\tilde \theta _{{f_i}}}^T{\tilde \theta _{{f_i}}} + \frac{1}{{2{\eta _g}}}{\tilde \theta _{{g_i}}}^T{\tilde \theta _{{g_i}}}\]

 its derivative is given by :

\begin{equation}
  \begin{aligned}
{\dot V_i} &= \left( {{U_z}{{\tilde \theta }_g}^T{\phi_g}{x}{u_{eq_i}} -{U_z}{w_g}(x){u_{eq_i}}- {U_1}\left({g_x} \right){u_{s_i}}} \right){e_\varepsilon } \\ 
& + \frac{1}{\eta _g}{\tilde \theta _g}^T{\dot \tilde \theta _g}- \gamma {e_\varepsilon }^2\,
 \end{aligned}
\end{equation}	

We take the following adaptation law:  ${\dot \theta _g}\left( t \right) =  - {\eta _g}{\varphi _g}\left( x \right){e_{{\varepsilon _x}}}{u_{eq}}{U_z}$

If the parameter $g_i$ is constant, so the adaptation law for the position controller is : ${\dot \theta _g}\left( t \right) =  - {\eta _g}{e_{{\varepsilon _x}}}{u_{eq_i}}{U_z}$ 
% and     ${U_1}{\tilde \theta _g}^T{\phi _g}{\text{(x)}}{u_{eq}} \to {U_z}{\tilde \theta _g}^T{u_{eq}}$    

\begin{equation}
  \label{lyap_prop3_dot}
 \begin{aligned}
{\dot V_i} &=  \left( {{U_z}{{\tilde \theta }_g}^T{\phi _g}{\text{(}}x{\text{)}}{u_{eq_i}} - {U_1}{w_g}(x){u_{eq_i}} - {U_z}\left( {g(x)} \right){u_s}} \right){e_s} \\
& - \gamma {e_\varepsilon }^2- {U_z}{\tilde \theta _g}^T{\varphi _g}\left( x \right){e_\varepsilon }{u_{{eq_i }}}\\
& \Rightarrow {\dot V_i} \leq  - \gamma {e_\varepsilon }^2\, + \left( { - {U_z}{w_g}(x){u_{eq_i}} - {U_z}\left( {{g_x}} \right){u_{s_i}}} \right){e_\varepsilon }
  \end{aligned}
\end{equation}			

To ensure that the equation Eqn.~(\ref{lyap_prop3_dot}) being less or equal to zero, we choose ${u_s}$ as follow:

\[{u_{{s_x}}} = \tfrac{g_i}{g_{0_i}}\left( {{W_{g_i}}(x)\left| {{u_{eq_i}}} \right|} \right){\text{sat}}(\frac{{{e_{{s_x}}}}}{\varepsilon })\]

We constate that : $ - \left( { + {w_g}(x){u_{eq_i}}} \right){e_\varepsilon } \leq \left( { + \left| {{w_g}(x){u_{eq_i}}} \right|} \right)\left| {{e_\varepsilon }} \right|$ so :

%\begin{equation}
\begin{multline*}
{\dot V_i} =   \left( - {U_1}{w_g}(x){u_{eq_i}}- {U_z}\left( \tfrac{g_i}{{g_0}_i} \right)\left( {{W_{g_5}}(x)|{u_{eq_i}}|} \right){\text{sat}}({e_s}/\varepsilon ) \right){e_\varepsilon } \\
 - \gamma {e_\varepsilon }^2 +\\
\end{multline*}
%\end{equation}

\begin{equation}
\begin{aligned}				
{\dot V_i} & \leq  - \gamma e_\varepsilon ^2 + (|{{U_z}{w_g}(x)u_{eq_i}} |)|{e_\varepsilon }|  \\
 &- {U_z}{e_\varepsilon }\left( {\tfrac{g_i}{{g_0}_i}\left( {{W_{g_i}}(x)\left| {{u_{eq_i}}} \right|} \right){\text{sat}}(\frac{e_s}{\varepsilon })} \right)
\end{aligned}
\end{equation}

With : $( \tfrac{g_i}{g_{0_i}} ) \geq 1 $ ,${U_z} \geq 0$ , $|{w_{f_i}}(x)| \leq {W_{f_i}}{(x)}$    and $|w_{g_i}(x)| \leq {W_{g_i}}(x)$  then $\forall {e_s} \ne 0$   we have   

  %- \gamma {e_\varepsilon }^2 + \left| {{U_1}{w_g}\left( x \right){u_{eq}}} \right|\left| {{e_\varepsilon }} \right|
 %- {U_1}{e_\varepsilon }{\text{sat}}\left( \frac{{e_s}}{\varepsilon } \right){W_g}\left( x \right)\left| {{u_{eq}}} \right|\\

\begin{equation}
\begin{aligned}     
&{\dot V_i} =\\
&- \gamma {e_\varepsilon }^2 + \left| {{U_z}{w_{g_i}}(x){u_{eq_i}}} \right|\left| {{e_\varepsilon }} \right| - {U_z}{e_\varepsilon }{\text{sat}}(\tfrac{e_s}{\varepsilon} ){W_{g_i}}(x)\left| {u_{eq_i}} \right| \\
& = {U_z}\left| {{e_\varepsilon }} \right|(\left| {w_{g_i}(x){u_{eq_i}}} \right| - {W_{g_i}}(x)\left| {u_{eq_i}} \right|) \leq 0 \\
{\dot V_i} &\leq  - \gamma {e_\varepsilon }^2
\end{aligned}
\end{equation}
since $\gamma {e_\varepsilon }^2 \geq 0 $ , this shows that $V_i$, is a nonincreasing function of time.
 
%By following the same procedure, the subsystem that govern $y$ can be controlled by:
       %${u_{{s_y}}} = \frac{1}{g_{06}}\left( { + {W_{g6}}(x)\left| {{u_{eq}}} \right|} \right){\text{sat}}(\frac{e_{s_y}}{\varepsilon })$   and   ${\dot \theta _{g6}}\left( t \right) =  - {\eta _{g6}}{e_{sy}}{u_{eq}}{{\text{U}}_1}$
			
The correspondent’s angles are given by the two following equations Eqn.~(\ref{equ_corr}):
\begin{equation}
\label{equ_corr}    
\begin{gathered}
  {\phi _d} = \arcsin ({u_x}\sin {\psi _d} - {u_y}\cos {\psi _d}) \hfill \\
  {\theta _d} = \arcsin (\frac{1}{{\cos \phi }}({u_x}\cos {\psi _d} - {u_y}\sin {\psi _d})) \hfill \\ 
\end{gathered} 
\end{equation}    

for $\psi \cong 0 $ we can approximate the Eqn.~(\ref{equ_corr}) using the following approximations:  
 $\cos {\psi _d} \approx 1\, \textnormal{and}\,\sin {\psi _d} \approx 0 \Rightarrow {\phi _d} = \arcsin ( - {u_y})$ and ${\theta _d} = \arcsin (\frac{1}{{\cos \phi }}{u_x})$

%%%%%%%%%%%%%%%%%%%%%%%%%%%%%%%%

\section{Results and discussions}
\label{res_discu}

To verify the reliability and robustness of the controller with simulations of tracking in the presence of different disturbance levels, considering the results obtained using the proposed control in Section ~\ref{sect_ctrl_quad} and considering the various trajectories where we use a sampling step of $0.01$ and ${W_{{f_i}}} = 0.1\,$ for  $i = \left\{ {\phi , \theta, \psi} \right\}$ and ${W_{g_j}} = 0.1$ for $\,j = \left\{ {\phi ,\theta,...,y,z} \right\}$ and  $ \gamma =10$ for attitude and altitude control, while $ \gamma =2$ and $K = [5 ,1]$ for position control. 

%%%%%%%%%%%%%%%%%%
\subsection{ Simulation with Disturbance}

\subsubsection{Attitude Control:} 
 this test is about applying our proposed control to quadrotor to study its behavior in addition to its performances.  Take the following trajectory: 

 \[\left\{ \begin{gathered}
  \,{\theta _d}(t) = sin(t)\,;{\varphi _d}(t) = sin(t + \pi )\,\,\,\, \hfill \\
  {\psi _d}(t) = 0.2\,\,\,\,\,\,{z_d}(t) = 1\,\,\,\,\, \hfill \\ 
\end{gathered}  \right.\forall t\]    
                   
\begin{figure} 
\begin{center}
%\centerline{\includegraphics[width=4.0in]{figure/fig10.pdf}}
\scalebox{0.56}{\includegraphics{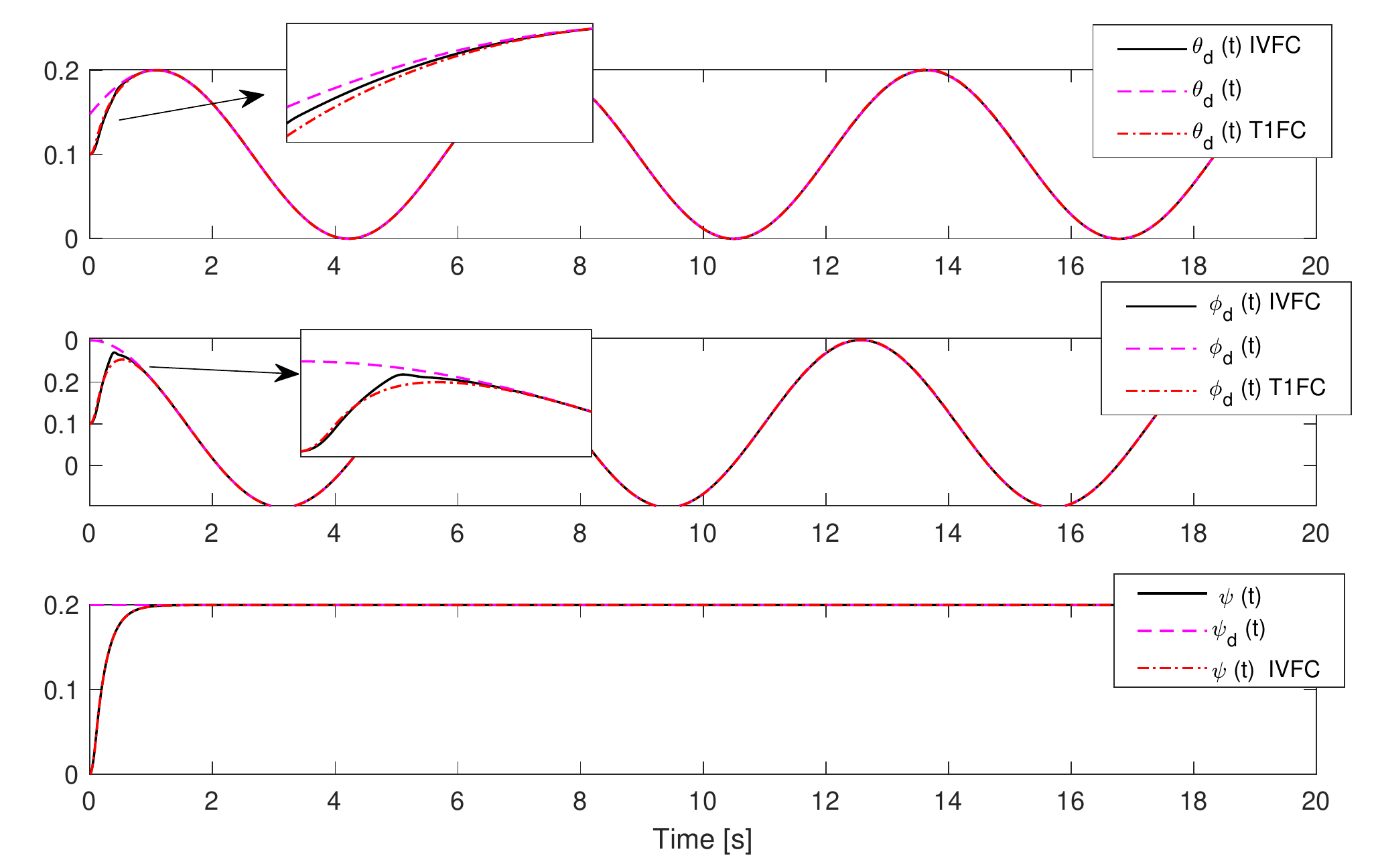}}
\caption{Reponses of system with a sinusoidal trajectory }
\label{fig10}
\end{center}
\end{figure}

\begin{figure} 
\begin{center}
%\centerline{\includegraphics[width=4.0in]{figure/fig11.pdf}}
\scalebox{0.56}{\includegraphics{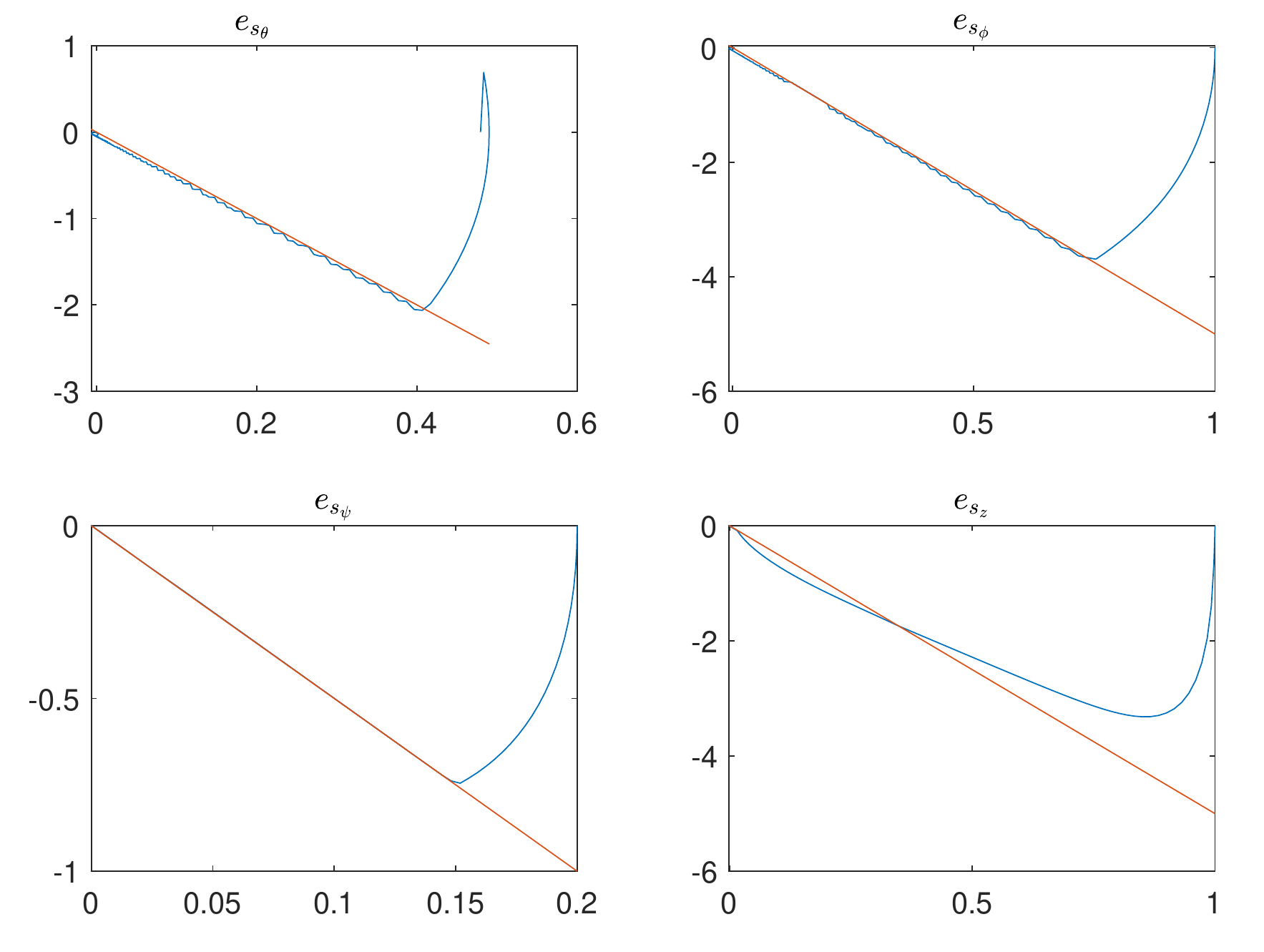}}
\caption{sliding errors $e_s$ }
\label{fig11}
\end{center}
\end{figure}

The desired and the real angles of output (in radians) are presented in Fig.~\ref{fig10}, and the sliding errors are presented in Fig.~\ref{fig11}.

%%%%%%%%%%%%%%%% begin figure %%%%%%%%%%%%%%%%%%%

%%%%%%%%%%%%%
For a complicated trajectory, we choose the following desired signal:

\begin{multline}
\left[\theta_d(t),\varphi_d(t),z_d(t),\psi_d(t) \right] = \left\{
 \begin{array}{l l}
       & \left[{0,0,1,0.2\,} \right]\,\,\,\,\,{\text{if}}\,t \in \left[ {10\,\,12} \right]\,\, \hfill \\
       & \left[{sin(t),sin(t + \pi ),1,0.2} \right]\,{\text{otherwise}}\, \hfill \\ 
    \end{array} \right.
 \end{multline}

\begin{figure} 
%\centerline{\includegraphics[width=4.0in]{figure/fig14.pdf}}
\begin{center}
\scalebox{0.55}{\includegraphics{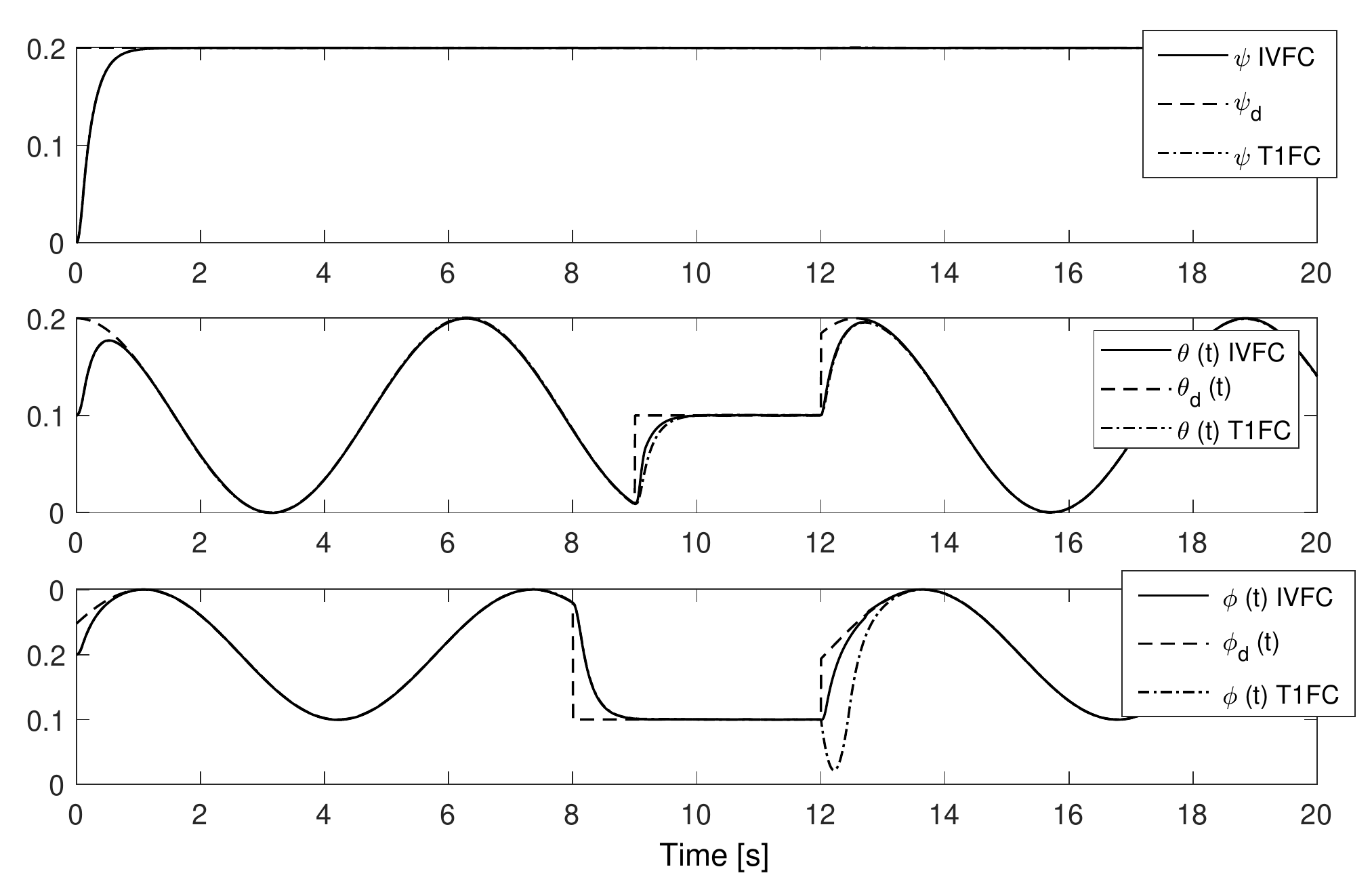}}
\caption{Responses of system with a sinusoidal trajectory }
\label{fig14}
\end{center}
\end{figure}

%\begin{equation}
%%\label{eq:t}
%\begin{aligned}
%\left[\theta_d(t),\varphi_d(t),z_d(t),\psi_d(t) \right] =\\
%\left\{
%\begin{gathered}
%\left[{0,0,1,0.2\,} \right]\,\,{\text{if}}\,t \in \left[ {10\,\,12} \right]\,\, \hfill \\
%\left[{sin(t),sin(t + \pi ),1,0.2} \right]\,{\text{otherwise}}\, \hfill \\ 
%\end{gathered}  \right.
  %%\hat f - f &= \theta _f^T\phi _f(x) - {\theta^{*}_{f}}^T\phi _f(x) - {w_f}(x) \\ 
	           %%&= \tilde \theta _f^T\phi _f(x) - w_f(x) \hfill \\ 
%\end{aligned}
%\end{equation}

desired angles and real angles of the system’s output (in radians) are shown in Fig.~\ref{fig14} .
%%%%%%%%%%%%%%%%%%%%%%%%%%%%%%%%%%%%%%%%

\begin{figure} 
%\centerline{\includegraphics[width=4.0in]{figure/fig15.png}}
\begin{center}
\scalebox{0.6}{\includegraphics{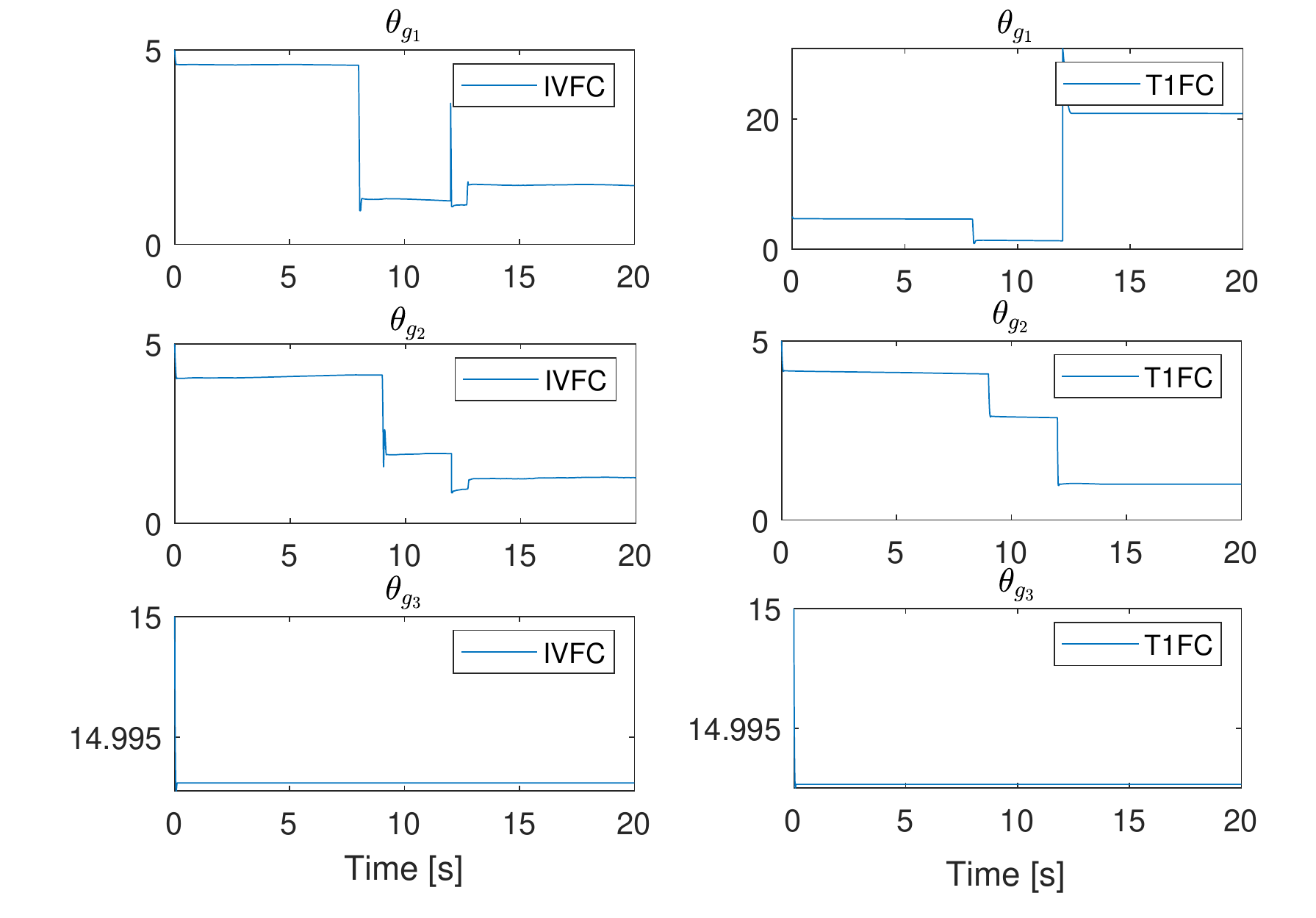}}
\caption{Adaptations of Parameters ${\theta _{g_\phi}},{\theta _{g_\theta}}$ and ${\theta _{g_\psi}}$  in function of time: a)T1FC b)IVFC }
\label{fig15}
\end{center}
\end{figure}

\begin{figure}
\begin{center} 
%\centerline{\includegraphics[width=4.0in]{figure/fig16.png}}
\scalebox{0.5}{\includegraphics{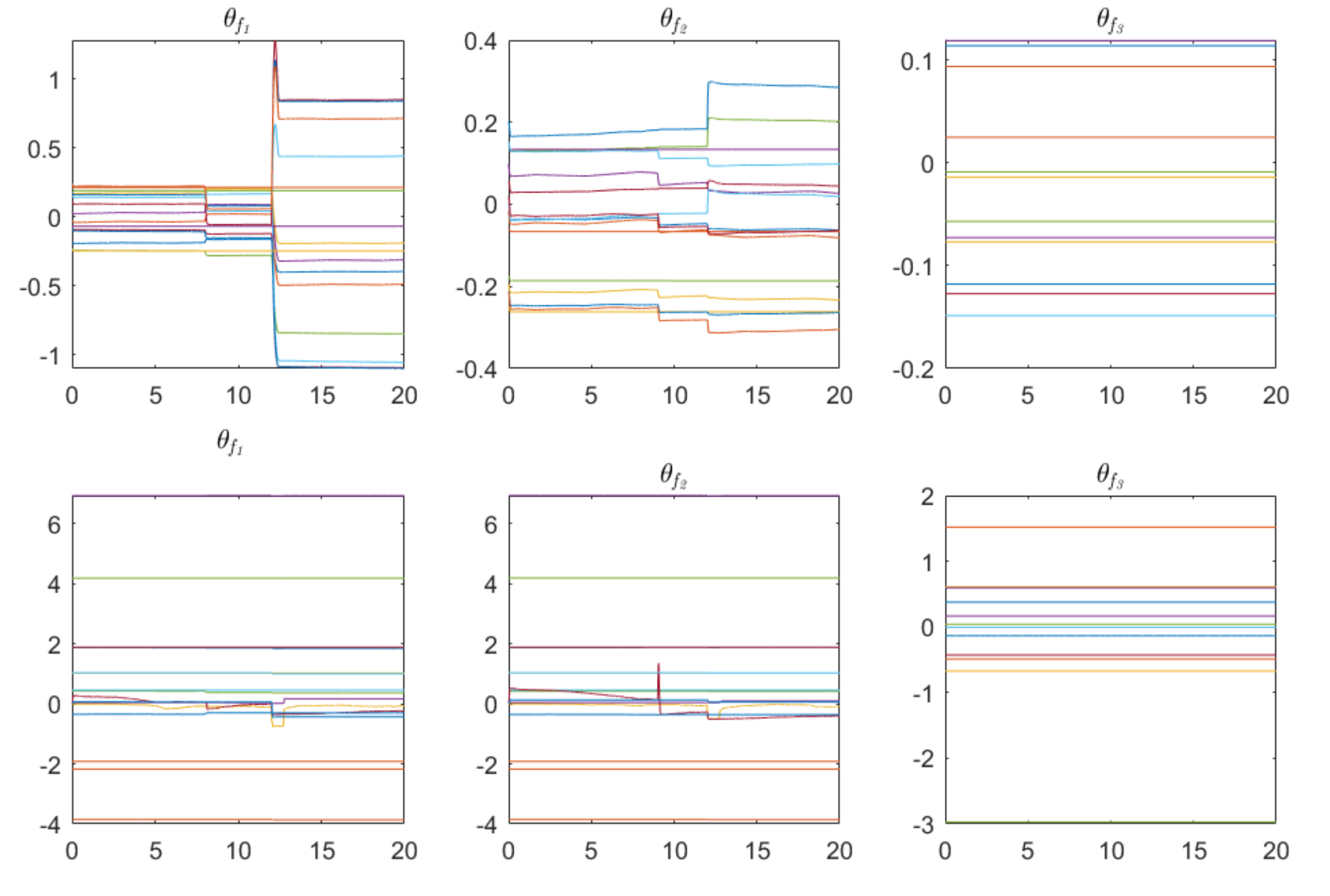}}
\caption{Parameters adaptation ${\theta _{f_\phi}},{\theta _{f_\theta}}$ and ${\theta _{f_\psi}}$  in function of time: a)T1FC b)IVFC }
\label{fig16}
\end{center}
\end{figure}

The figures Fig.~\ref{fig15}  and  Fig.~\ref{fig16} depict the evolution of the fuzzy model's adaptation parameters. It is clear that parameters converge rapidly to their stationary values. Those peaks are due to system perturbations.

%%%%%%%%%%%%%%%%%%%%%%%% figure
%
%
%\end{figure}%
%\begin{figure}[ht]\ContinuedFloat
    %\centering
    %\begin{subfigure}{0.6\textwidth}
        %\includegraphics[width=\textwidth]{figure/fig14.pdf}
        %\subcaption{$Q^{*}$ values for arm 4}
        %\label{fig:arm3}
    %\end{subfigure}
%
    %\begin{subfigure}{0.6\textwidth}
        %\includegraphics[width=\textwidth]{figure/fig15.pdf}
        %\subcaption{$Q^{*}$ values for arm 5}
        %\label{fig:arm4}
    %\end{subfigure}
		%
		%
		    %\begin{subfigure}{0.6\textwidth}
        %\includegraphics[width=\textwidth]{figure/fig16.pdf}
        %\subcaption{$Q^{*}$ values for arm 5}
        %\label{fig:arm4}
    %\end{subfigure}
 %
		%
%\end{figure}
%

%%%%%%%%%%%%%%%%%%%%%

We note that the system met the desired reference. The error of tracking in Fig.~\ref{fig14} and the Table~\ref{table_comp1} can be explained by the error between the estimated model and the real model. Anyway, the effect of this error is too small, and we note that the error obtained with this approach is minimal.
%%%%% table 3

\begin{table}[t]
\caption{Comparison of MSE between T1 and IVFC}
\begin{center}
\label{table_comp1}
\begin{tabular}{c l l}
& \\ % put some space after the caption
\hline
MSE & IVFC & T1FC \\
\hline
$\tilde \theta $ &  22.5090	 & 23.0242\\
$\tilde \beta $ & 44.9816 &  81.8549\\
$\tilde \psi $ &  1.0554   & 1.0587\\
\hline
\end{tabular}
\end{center}
\end{table}

%%%%%%%%%%%%%%%

\subsubsection{Position Control:} 
we consider the following trajectory:
$$
  {x_d}(t) = sin(t)\,,{y_d}(t) = sin(t + \pi ) \, ,{z_d}(t) = 1\,\,,\,\,{\psi _d}(t) = 0.2
$$

\begin{figure} 
\begin{center}
%\centerline{\includegraphics[width=5in]{figure/fig17.png}}
\scalebox{0.6}{\includegraphics{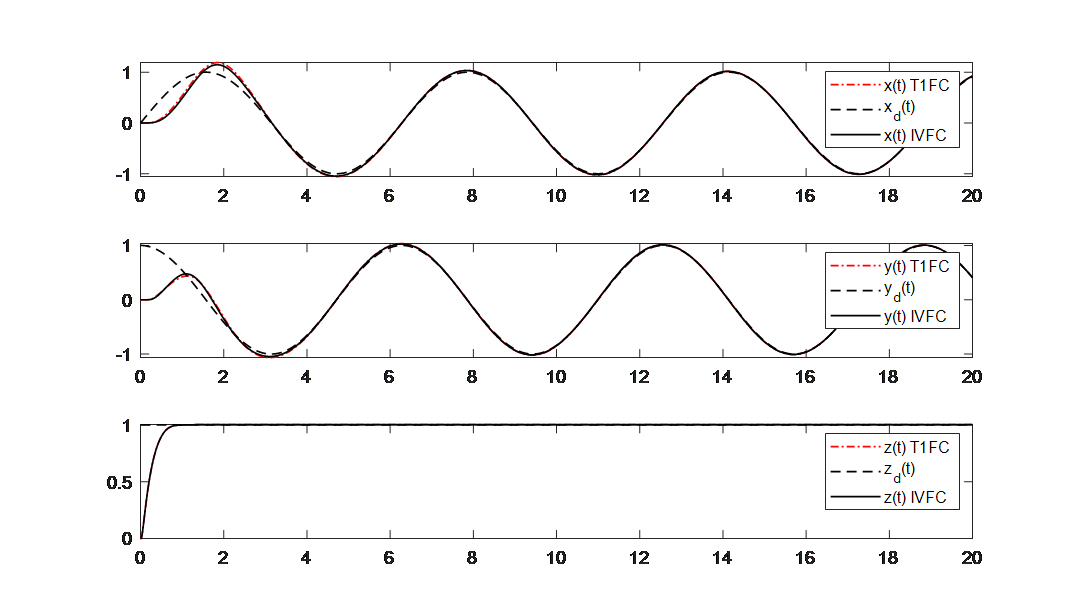}}
\caption{response of system with sinusoidal trajectory  }
\label{fig17}
\end{center}
\end{figure}

\begin{figure} 
\begin{center}
%\centerline{\includegraphics[width=5in]{figure/fig18.pdf}}
\scalebox{0.5}{\includegraphics{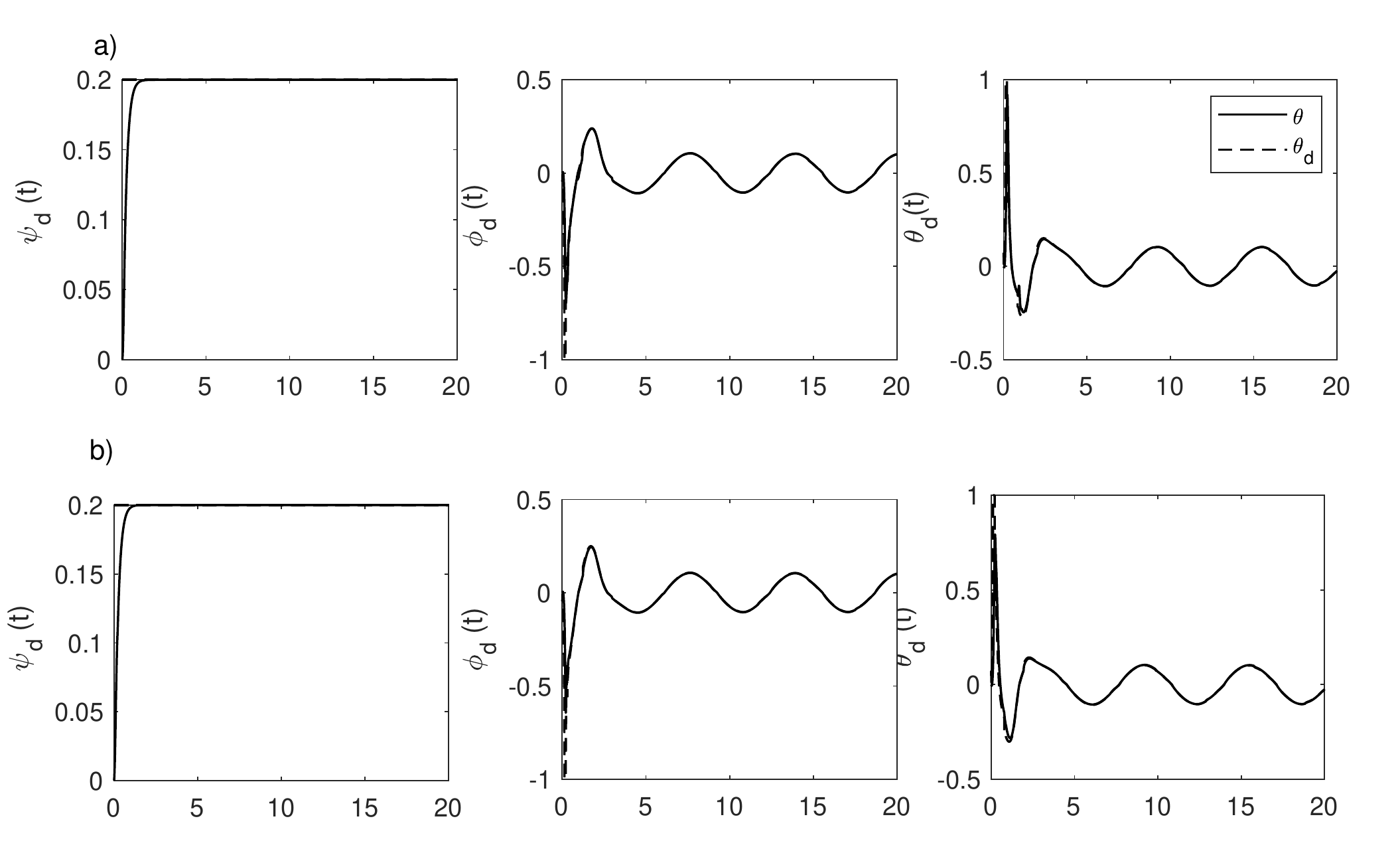}}
\caption{correspondent’s Angles to responses of position a) T1FC b) IVFC }
\label{fig18}
\end{center}
\end{figure}

%\left\{
%\begin{gathered}
%
%\end{gathered} 
 %\right.
%$$                             
In the absence of any parametric variation in the case of position tracking, the results of simulation are almost identical and sufficiently acceptable Fig.~\ref{fig17}. Figure Fig.~\ref{fig18} depicts the angles that correspond to position responses. 

%%%%%%%%%%%%%%%%%%%%%%%% figure

%%%%%%%%%%%%%%%%%%

\subsection{Simulation with Perturbation}

\subsubsection{Attitude Control:} 

This test is about the application of the control law to a quadrotor with perturbations in the range of time [12s,14s].  This test is carried out by varying  $\delta{I}$  and $\delta{J_r}$  around the values of the rotor's inertia ${J_r}$ and the drone's inertia $(I_{xx},I_{yy},I_{zz})$. The variation can be written as ${\hat J_r} = {J_r} + \delta {J_r}$   and   $ \hat I = I + \delta I $.

\begin{figure} 
\begin{center}
%\centerline{\includegraphics[width=5in]{figure/fig19.png}}
\scalebox{0.6}{\includegraphics{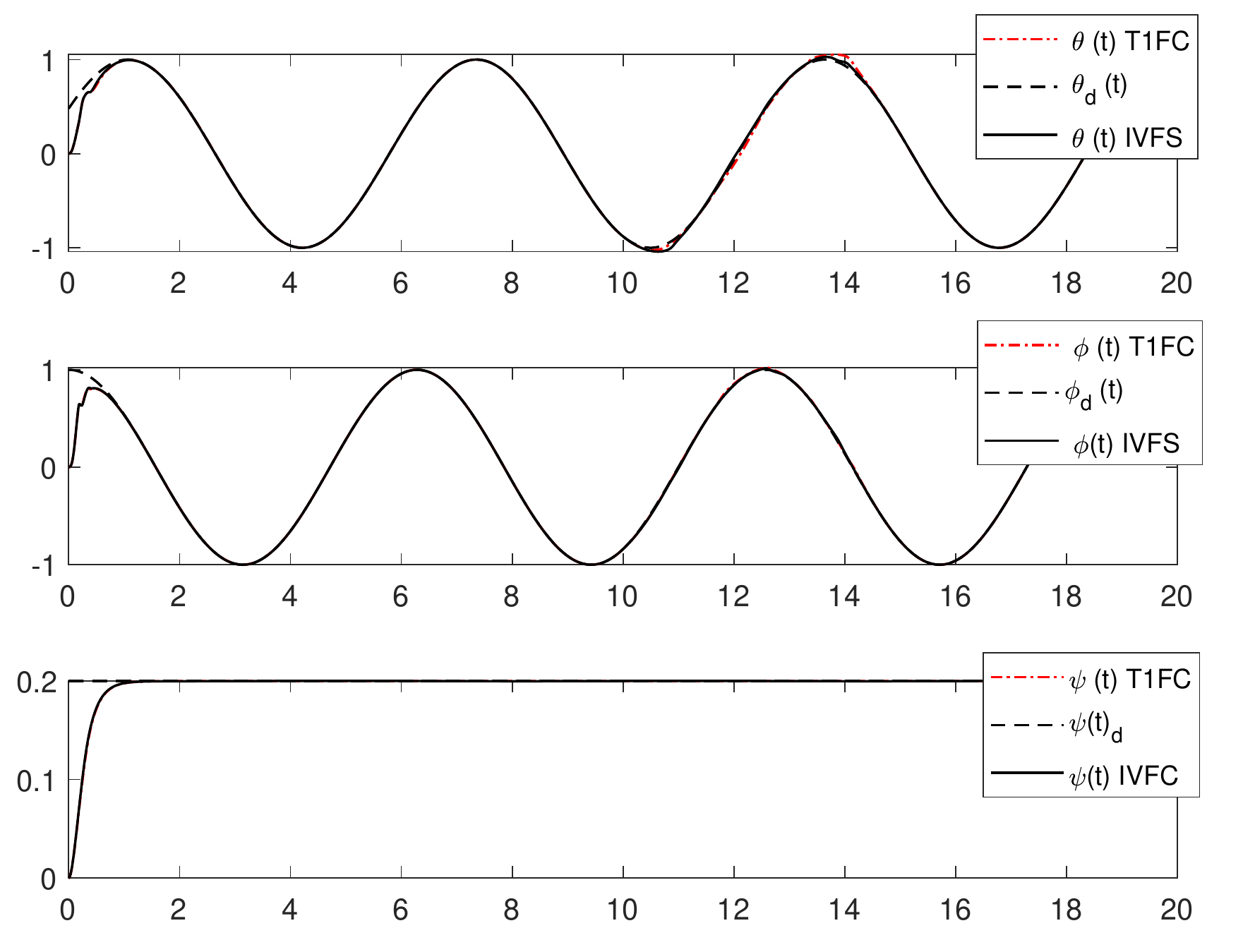}}
\caption{Adaptative Sliding mode control in presence of disturbance  $\delta {J_r} = 10\% {J_r}$ }
\label{fig19}
\end{center}
\end{figure}

\begin{figure} 
\begin{center}
%\centerline{\includegraphics[width=5in]{figure/fig20.png}}
\scalebox{0.6}{\includegraphics{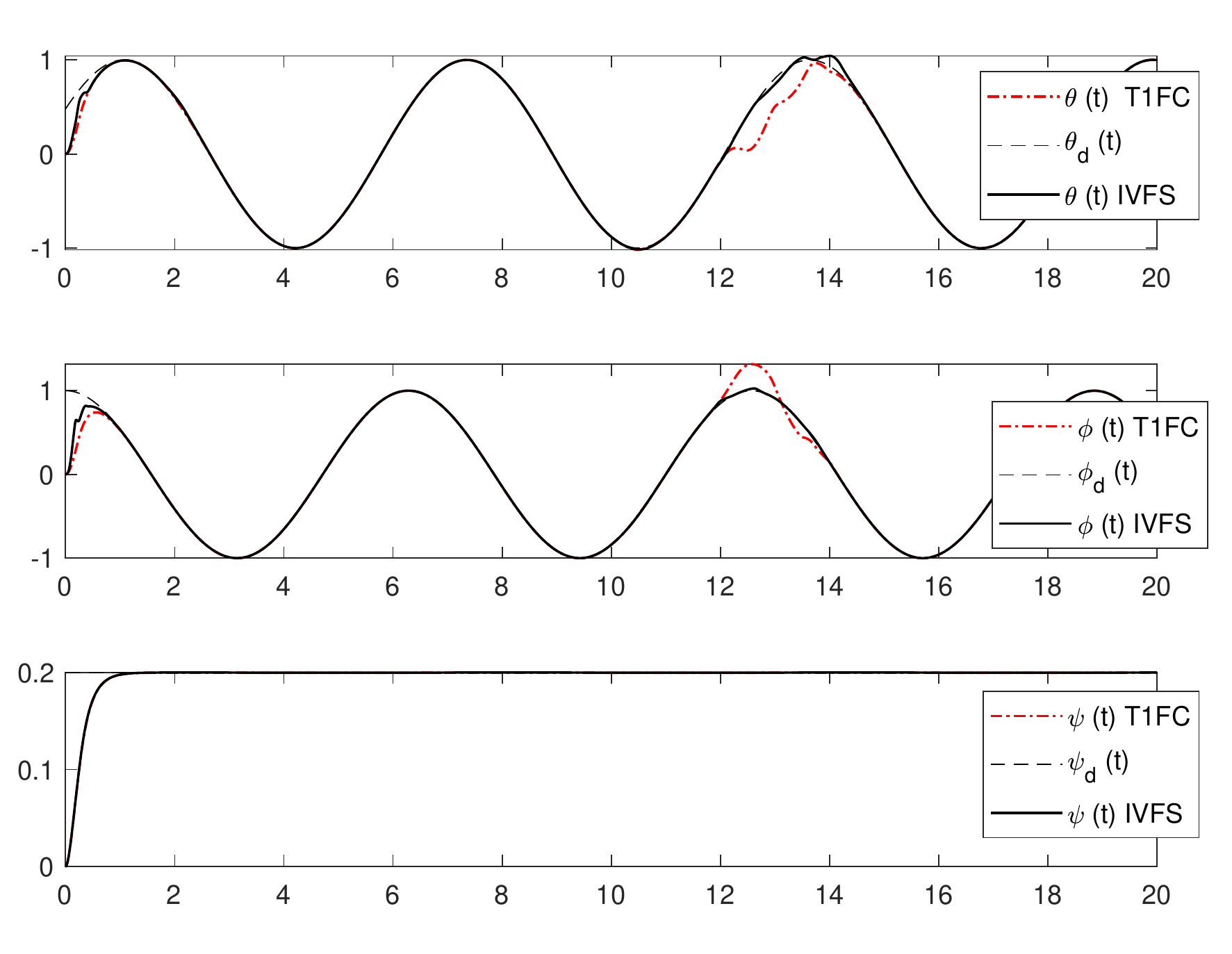}}
\caption{Adaptative Sliding mode control in presence of disturbance  $\delta {J_r} = 15\% {J_r}$ }
\label{fig20}
\end{center}
\end{figure}

\begin{figure}
\begin{center}
%\centerline{\includegraphics[width=5.0in]{figure/fig21.png}}
\scalebox{0.55}{\includegraphics{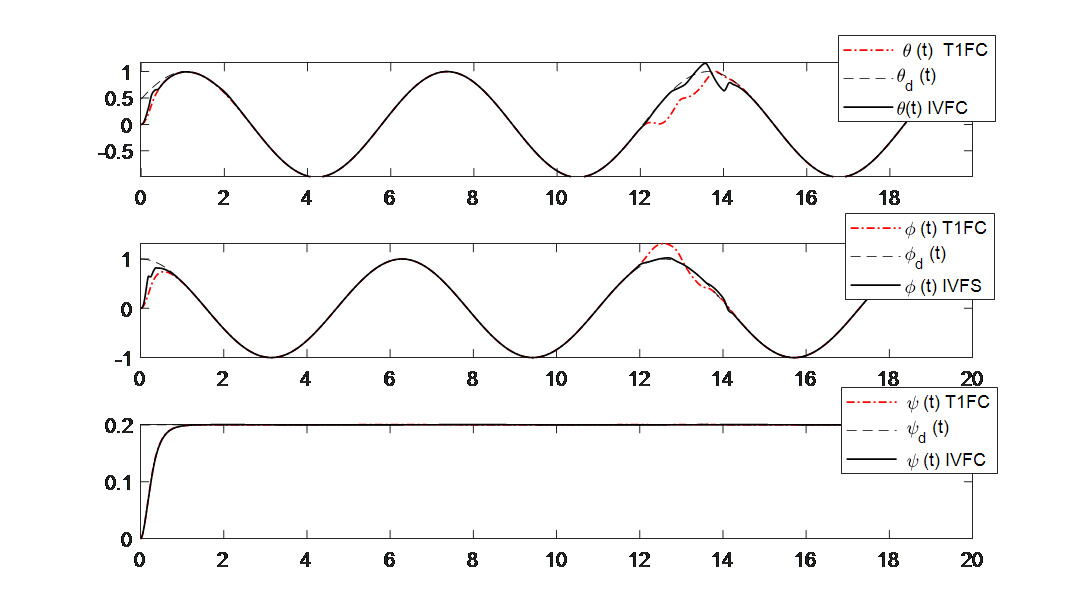}}
\caption{ Adaptive sliding mode controller in presence of perturbations$\delta {J_r} = 20\% {J_r}$   }
\label{fig21}
\end{center}
\end{figure}

 Figures (Fig.~\ref{fig19} ~\ref{fig20} and ~\ref{fig21}) show the results of tracking for the three tests mentioned above. In the case where we introduce the model disturbance, we note a deviation from the desired consign.

%%%%%%%%%%%%%%%%%%

%%%%%%%%%%%%%%%%%%%%%%%%%%%%%%
 Note that in all simulations, the added perturbations are quickly compensated by the controller IVFC proposed with respect to the conventional type-1 controller. Note also that in Table~\ref{table_comp11} the rejection of perturbation is more effective, especially when the system has a high level of disturbance.

	%Test 3	Test 2	Test 1
	%IVFC	T1FC	IVFC	T1FC	IVFC	T1FC
%MSE 
%7.4922	13.2162	7.5905	9.6931	8.0607	9.1189
%MSE 
%17.5483	49.8297	17.3102	32.5082	17.2950	31.0333
%MSE 
%1.0554	1.0561	1.0554	1.0554	1.0554	1.0537

%\begin{table}
%\caption{Comparison of MSE between T1 and IVFC}
%\begin{center}
%\label{table_comp11}
%\resizebox{\columnwidth}{!}{%
%\begin{tabular}{|c|l|l|l|l|l|l|}
 %& \multicolumn{2}{c}{test 1} & \multicolumn{2}{c}{test 2} & \multicolumn{2}{c}{test 3} \\ \hline
%\multicolumn{1}{c}{MSE} & IVFC	& T1FC    &	IVFC  &	T1FC  &	IVFC  &	T1FC\\ \hline
%$\tilde \theta $        &7.492 &	13.216	&7.590	&9.693	&8.060	&9.118\\ \hline
%$\tilde \beta $         & 17.548	&49.829	&17.310&	32.508&	17.295&	31.033\\ \hline
%$\tilde \psi $          & 1.055	&1.056	&1.055  	&1.055  &1.055	&1.053\\ \hline
%\hline      
%\end{tabular}
%}
%\end{center}
%\end{table}

%%%%%%%%%%%%%%%%%%%%%%%

\begin{table}
\caption{Comparison of MSE between T1 and IVFC}
\begin{center}
\label{table_comp11}
%\resizebox{\columnwidth}{!}{%
\begin{tabular}{cllllll}
\hline
 & \multicolumn{2}{c}{test 1} & \multicolumn{2}{c}{test 2} & \multicolumn{2}{c}{test 3} \\
\hline
\multicolumn{1}{c}{MSE} & IVFC	& T1FC    &	IVFC  &	T1FC  &	IVFC  &	T1FC\\
\hline
$\tilde \theta $        &7.492 &	13.216	&7.590	&9.693	&8.060	&9.118\\
$\tilde \beta $         & 17.548	&49.829	&17.310&	32.508&	17.295&	31.033\\
$\tilde \psi $          & 1.055	&1.056	&1.055  	&1.055  &1.055	&1.053\\
\hline      
\end{tabular}
%}
\end{center}
\end{table}

%%%%%%%%%%%%%%%%%%%%

\begin{figure} 
\begin{center}
%\centerline{\includegraphics[width=5.0in]{figure/fig22.png}}
%\centerline{\includegraphics[width=5.0in]{figure/fig22.pdf}}
\scalebox{0.5}{\includegraphics{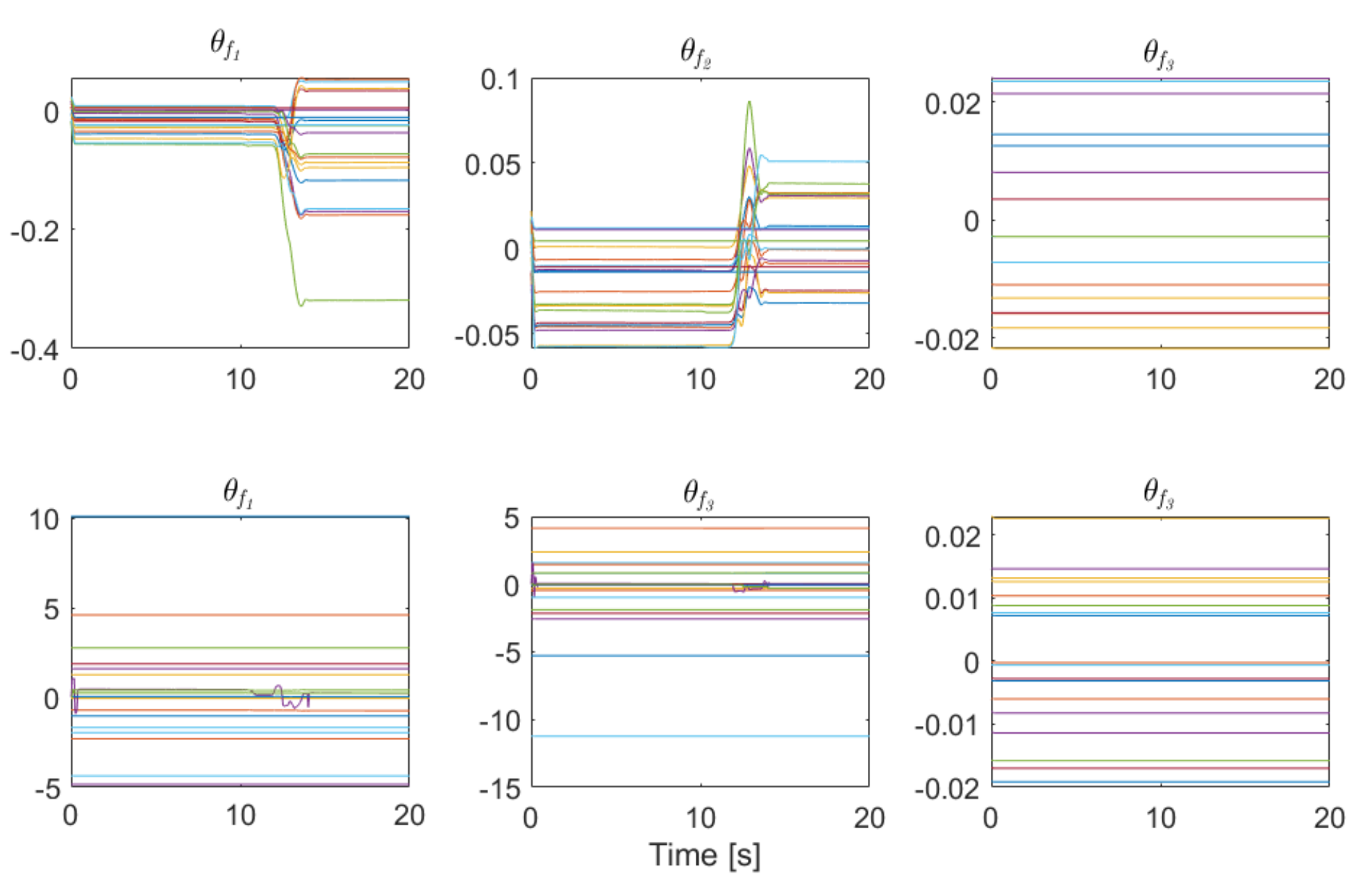}}
\caption{Evolution of Adaptations parameters of  ${f_i}(x)$ : a) T1FC b) IVFC}
\label{fig22}
\end{center}
\end{figure}
%%%%%%%%%%%%%%%%%%%%%

\subsubsection{Position Control:}

\begin{figure} 
\begin{center}
%\centerline{\includegraphics[width=5.0in]{figure/fig23.pdf}}
\scalebox{0.55}{\includegraphics{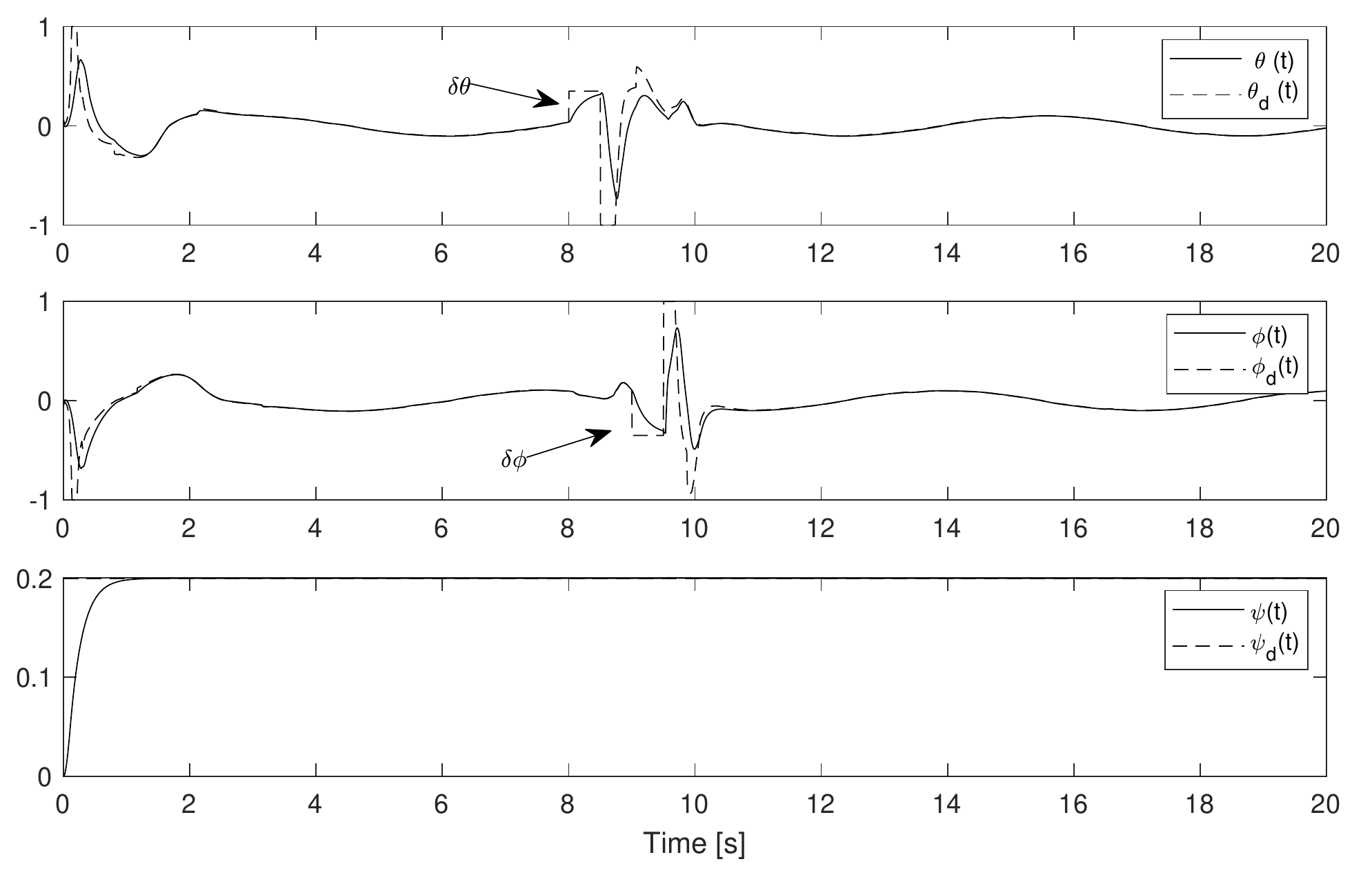}}
\caption{Angles correspondent with perturbation  $\delta \varphi  = \delta \theta  = 0.25rad $ }
\label{fig23}
\end{center}
\end{figure}

\begin{figure} 
\begin{center}
%\centerline{\includegraphics[width=5.0in]{figure/fig24.png}}
\scalebox{0.55}{\includegraphics{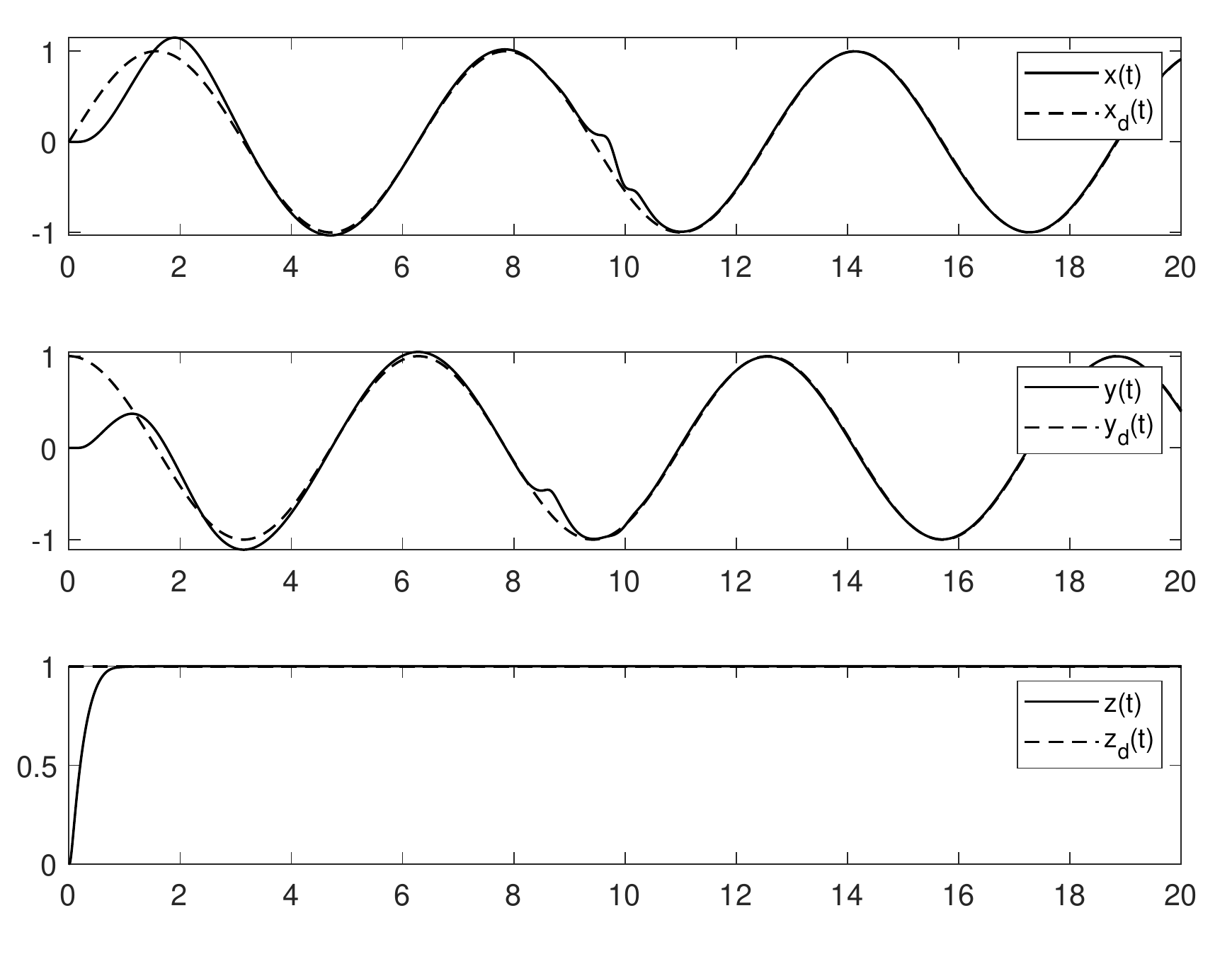}}
\caption{Response of system using sliding mode adaptative fuzzy model based}
\label{fig24}
\end{center}
\end{figure}

In this simulation, the model is submitted to an external disturbance on the angles ${\varphi _d}$ and ${\theta _d}$ that attack the position controller. This additive perturbation is chosen to be $\delta \varphi = 0.25rad$ in the range [8s 8.5s] and $\delta \theta = 0.25rad$ in the range [9s 9.5s], as shown in the  Fig.~\ref{fig23}. The tracking response of position is shown in Fig.~\ref{fig24} .

%%%%%%%%%%%%figures 

%%%%%%%%%%%%%%%%%%%%

The results of Fig.~\ref{fig24} and the MSE Table~\ref{table_comp2} show the efficiencies and  robustness of the proposed control.

%%%%%%%%%%%%%%%%%%%%%%  tab 5 %%%%%%%%%%%%%%%
\begin{table}[t]
\caption{Comparison of MSE between T1 and IVFC}
\begin{center}
\label{table_comp2}
\begin{tabular}{c l l}
& & \\ % put some space after the caption
\hline
MSE & T1FC & IVFC \\
\hline
$\tilde x$ &  46.6442 & 39.7494\\
$\tilde y$ & 127.3148 &  123.8104 \\
$\tilde z$ &  30.7955   & 30.7955\\
\hline
\end{tabular}
\end{center}
\end{table}
%%%%%%%%

\section{CONCLUSION }

In this paper, we presented an algorithm to build an IVFM fuzzy model from input-output data for a nonlinear system. The proposed method for constructing the Takagi-Sugeno (TS) type-2 fuzzy model, based on input-output data of the identified dynamics, is constructed by identifying the structure using fuzzy clustering, then, the envelope detection, and finally the identification of parameters. The proposed method is used to model the unknown parts of the quadrotor dynamics, and the model is then used to design a fuzzy sliding mode adaptive controller in which the unknown dynamics are approximated by an IVFM as an initial model, and the model is then adjusted online using Lyapunov theory-based law. The compared results are presented to demonstrate the effectiveness of the proposed method. Many cases are considered for the sake of showing and testing the performance of tracking and robustness. Finally, the results show that the tracking error has converged despite internal and external disturbances.

%%%%%%%%%%%%%%%%%%%%%%%%%%%%%
%%%%%%%%%%%%%%%%%%%%%%%%%%%%%%

\bibliographystyle{unsrtnat}
\bibliography{template}  %%% Uncomment this line and comment out the ``thebibliography'' section below to use the external .bib file (using bibtex) .





% \end{thebibliography}

\end{document}